\pdfoutput=1
\documentclass[aps,nofootinbib,prd,preprintnumbers]{revtex4}

\usepackage{wrapfig,amsmath,hyperref,graphicx,epsfig,bm,color,verbatim,slashed,mathtools,xparse,adjustbox,dcolumn,float,relsize,tikz,cancel}

\usepackage{subcaption}
\usepackage[english]{babel}
\usepackage[utf8]{inputenc}
\begin{document}
	
	\title{Is it possible to explain the muon and electron $g-2$ in a $Z^{\prime }$ model?}
	\author{A. E. C\'{a}rcamo Hern\'{a}ndez${}^{a}$}
	\email{antonio.carcamo@usm.cl}
	\author{S. F. King${}^{b}$}
	\email{king@soton.ac.uk}
	\author{H. Lee${}^{b}$}
	\email{hl2n18@soton.ac.uk}
	\author{S. J. Rowley${}^{b}$}
	\email{s.rowley@soton.ac.uk}
	\affiliation{$^{{a}}$Universidad T\'{e}cnica Federico Santa Mar\'{\i}a and Centro Cient%
		\'{\i}fico-Tecnol\'{o}gico de Valpara\'{\i}so, \\
		Casilla 110-V, Valpara\'{\i}so, Chile, $^{{b}}$School of Physics and
		Astronomy, University of Southampton,\\
		SO17 1BJ Southampton, United Kingdom }
	\date{\today }
	
\begin{abstract}
In order to address this question, we consider a simple renormalisable and gauge invariant
model in which the $Z'$ only has couplings to the electron and muon and their associated neutrinos, 
arising from mixing with a heavy vector-like fourth family of leptons. 
Within this model we discuss the contributions to the electron and muon anomalous magnetic moments from $Z'$ exchange, 
subject to the constraints from $\mu \rightarrow e \gamma$ and neutrino trident production.
Using analytic and numerical arguments, we find that such a $Z'$ model can account for either the electron or the muon $g-2$ anomalies, but not both, while remaining consistent with the experimental constraints from $\mu \rightarrow e \gamma$ and neutrino trident production.\\[1.5ex]
\footnotesize
DOI:\href{https://doi.org/10.1103/PhysRevD.101.115016}{10.1103/PhysRevD.101.115016}
\normalsize
\end{abstract}

\maketitle

\section{Introduction}

The Standard Model (SM) provides an excellent explanation of all experimental data, apart from neutrino mass and lepton mixing.
Yet there are a few possible anomalies in the flavour sector that may indicate new physics beyond the SM.
For example, recently, there have been hints of universality violation in the charged lepton sector from $B\rightarrow K^{(\ast )}l^{+}l^{-}$ decays by the LHCb collaboration\cite{Descotes-Genon:2013wba,Altmannshofer:2013foa,Ghosh:2014awa}. Specifically, the $R_{K}$ \cite{Aaij:2019wad} and $R_{K^{\ast }}$ \cite{Bifani} ratios of $\mu ^{+}\mu ^{-}$ to $e^{+}e^{-}$ final states in the $B\rightarrow K^{(\ast)}l^{+}l^{-}$ decays are observed to be about $70\%$ of their expected values with a roughly $2.5\sigma$ deviation from the Standard Model (SM) in each channel. Following the recent measurement of $R_{K^{\ast }}$ \cite{Bifani}, a number of phenomenological analyses have been presented\cite{Hiller:2017bzc,Ciuchini:2017mik,Geng:2017svp,Capdevila:2017bsm,Ghosh:2017ber,Ciuchini:2019usw,Bardhan:2017xcc} that favour a new effective field theory (EFT) physics operator of the $C_{9\mu }^{NP}=-C_{10\mu }^{NP}$ form\cite{Glashow:2014iga,DAmico:2017mtc,Aebischer:2019mlg}. The most recent global fit of this operator combination yields ${C_9 = (34.0\text{TeV}) ^{-2}}$\cite{Aebischer:2019mlg}, though other well-motivated solutions are also possible\cite{Descotes-Genon:2015uva}.

In previous works\cite{Glashow:2014iga}, it has been suggested that such observations of charged lepton universality violation (CLUV) must be accompanied by charged lepton flavour violation (CLFV) such as $\mu\rightarrow e \gamma$ in the same sector, however, such a link cannot be established in a model-independent way because the low-energy effective operators for each class of processes are different. Nevertheless, in concrete models the connection is often manifest. This motivates studies of specific models. For example, studies of CLFV in B-decays using generic $Z^{\prime }$ models (published before the $R_{K^{\ast }}$ measurement but compatible with it) are provided in Ref. \cite{Crivellin:2015era}. A concise review of BSM scenarios that aim to explain CLUV and possible connections to dark matter is provided in Ref. \cite{Vicente:2018xbv}. Other theoretical explanations for universality violation in the lepton sector are discussed in Refs. \cite{Glashow:2014iga,Crivellin:2015era,Crivellin:2015lwa,Bonilla:2017lsq,Assad:2017iib, King:2017anf,Romao:2017qnu,Antusch:2017tud,Ko:2017quv,King:2018fcg,Falkowski:2018dsl,Barman:2018jhz,CarcamoHernandez:2018aon,deMedeirosVarzielas:2018bcy,Rocha-Moran:2018jzu,Hu:2018veh,Carena:2018cow,Babu:2018vrl,Allanach:2018lvl,Fornal:2018dqn,Aydemir:2018cbb,CarcamoHernandez:2019cbd,CarcamoHernandez:2019xkb,Aydemir:2019ynb}.

Independently of these anomalies, for some time now, it has been known that the experimentally measured anomalous magnetic moments 
g-2 of both the muon and electron each observe a discrepancy of a few standard deviations with respect to the Standard Model predictions. The longstanding non-compliance of the muon g-2 with the SM was first observed by the Brookhaven E821 experiment at BNL~\cite{Bennett:2006fi}. The electron g-2 has more recently revealed a discrepancy with the SM, following an accurate measurement 
of the fine structure constant~\cite{Parker:2018vye}.
However the different magnitude and opposite signs of the electron and muon g-2 deviations makes it difficult to explain both of these anomalies in any model, which also satisfies the constraints of CLFV, with all existing simultaneous explanations involving 
new scalars~\cite{Crivellin:2018qmi,Giudice:2012ms,Davoudiasl:2018fbb,Liu:2018xkx,Bauer:2019gfk,Han:2018znu,Dutta:2018fge,Badziak:2019gaf,Endo:2019bcj}, or conformal extended technicolour~\cite{Appelquist:2004mn}. We know of no study which discusses both anomalies in a $Z'$ model.
One possible reason is that the CLFV process 
$\mu\rightarrow e \gamma$, which would be concrete of BSM physics in the charged fermion sector,
is very constraining. Neutrino phenomena do give rise to CLFV but in the most minimal extensions this would occur at a very low rate in the charged sector, making it practically unobservable. Given the considerable resources committed to looking for CLFV, it is crucial to study relevant, well-motivated BSM scenarios which allow for CLFV at potentially observable rates. For example, such decays can be enhanced by several orders of magnitude if one considers extensions of the SM with an extra $U(1)'$ gauge symmetry spontaneously broken at the TeV scale. To summarise, although such extensions are able to successfully accommodate the experimental value of the muon magnetic moment\cite{Falkowski:2018dsl,Allanach:2015gkd,Raby:2017igl,CarcamoHernandez:2019xkb,Kawamura:2019rth}, we know of no study of a $Z'$ model which discusses both the electron and muon magnetic moments, including the constraints from $\mu\rightarrow e \gamma$.

In this work, we ask the question: is it possible to explain the anomalous muon and electron $g-2$ in a $Z^{\prime }$ model?
It is difficult to answer this question in general, since there are many possible $Z'$ models.
However it is possible to consider a model in which the $Z'$ only has couplings to the electron and muon and their associated 
neutrinos, arising from mixing with a vector-like fourth family of leptons, thereby 
eliminating the quark couplings and allowing us to focus on 
the connection between CLUV, CLFV and the electron and muon g-2 anomalies.
Such a renormalisable and gauge invariant 
model is possible within a $U(1)'$ gauge extension of the SM augmented by a fourth, vector-like family of fermions and right-handed neutrinos as proposed in \cite{King:2017anf}. In the fermiophobic version of this model~\cite{King:2017anf}, only the fourth family carry $U(1)'$ charges, with the three chiral families not coupling to the $Z'$ in the absence of mixing. Then one can switch on mixing between the first 
and second family of charged leptons and the fourth family, allowing controlled couplings of the $Z'$ to only the electron and muon
(and fourth family leptons) of the kind
we desire. Such a model allows charged lepton universality violation (CLUV) at tree-level with CLFV and contributions to 
the electron and muon magnetic moments at loop level. 
Within such a model we attempt to explain 
the anomalous magnetic moments of both the muon and electron within the relevant parameter space of the model,
while satisfying the constraints of $BR(\mu\rightarrow e\gamma)$ and neutrino trident production.
Using both analytic and numerical arguments, 
we find that it is not possible to simultaneously explain the electron and muon g-2 results consistent with these constraints.

The remainder of this article is organised as follows; in Section \ref{sec:model} we outline the renormalisable and gauge invariant 
fermiophobic model in which the $Z'$ couples only to a vector-like fourth family. 
In Section \ref{sec:LFV_in_this_model}, we show how it is possible to switch on the couplings of 
the $Z'$ to the electron and muon and their associated 
neutrinos, thereby eliminating all unnecessary couplings and allowing us to focus on 
the connection between CLUV, CLFV and the electron and muon g-2 anomalies. A simplified analytical analysis of the CLFV and the electron and muon g-2 anomalies in the fermiophobic $Z'$ Model is presented in Section \ref{sec:analytics}. In Section \ref{sec:statistical_analysis} we analyse the parameter space numerically, presenting detailed predictions for each of the examined leptonic phenomena.
Section \ref{sec:conclusion} concludes the paper.

\section{The Fermiophobic $Z'$ Model}
\label{sec:model}
Consider an extension of the SM with a $U(1)'$ gauge symmetry, where fermion content is expanded by right-handed neutrinos and a fourth, vector-like family. The scalar sector is augmented by gauge singlet fields with non-trivial charge assignments under the new symmetry. The basic framework for such a theory was defined in  \cite{King:2017anf}.
Henceforth we consider the case where the 
SM fermions in our model are uncharged under the additional symmetry, whereas the vector-like fermions are charged under this symmetry, corresponding to so called ``fermiophobic $Z'$'' model considered in \cite{King:2017anf}. 
The field content and charge assignments are given in Table \ref{tab:model_content}.
Note that such a theory is anomaly free; left- and right-handed fields of the vector-like fermion family have identical charges under $U(1)'$, and hence chiral anomalies necessarily cancel. 
\begin{table}[th]
	\centering\renewcommand{\arraystretch}{1.3} 
	\begin{tabular}{|c||c|c|c|c|c|c|c|c|c|c|c|c|c|c|c|c|c|c|c|c|}
		\hline
		Field & $Q_{iL}$ & $u_{iR}$ & $d_{iR}$ & $L_{iL}$ & $e_{iR}$ & $\nu_{iR}$ & 
		$H$ & $Q_{4L}$ & $\widetilde{Q}_{4R}$ & $\widetilde{u}_{4L}$ & $u_{4R}$ & $%
		\widetilde{d}_{4L}$ & $d_{4R}$ & $L_{4L}$ & $\widetilde{L}_{4R}$ & $%
		\widetilde{E}_{4L}$ & $E_{4R}$ & $\nu _{4R}$ & $\widetilde{\nu }_{4L}$ & $%
		\phi _{f}$ \\ \hline\hline
		$SU(3)_{c}$ & $\mathbf{3}$ & $\mathbf{3}$ & $\mathbf{3}$ & $\mathbf{1}$ & $%
		\mathbf{1}$ & $\mathbf{1}$ & $\mathbf{1}$ & $\mathbf{3}$ & $\mathbf{3}$ & $%
		\mathbf{3}$ & $\mathbf{3}$ & $\mathbf{3}$ & $\mathbf{3}$ & $\mathbf{1}$ & $%
		\mathbf{1}$ & $\mathbf{1}$ & $\mathbf{1}$ & $\mathbf{1}$ & $\mathbf{1}$ & $%
		\mathbf{1}$ \\ \hline
		$SU(2)_{L}$ & $\mathbf{2}$ & $\mathbf{1}$ & $\mathbf{1}$ & $\mathbf{2}$ & $%
		\mathbf{1}$ & $\mathbf{1}$ & $\mathbf{2}$ & $\mathbf{2}$ & $\mathbf{2}$ & $%
		\mathbf{1}$ & $\mathbf{1}$ & $\mathbf{1}$ & $\mathbf{1}$ & $\mathbf{2}$ & $%
		\mathbf{2}$ & $\mathbf{1}$ & $\mathbf{1}$ & $\mathbf{1}$ & $\mathbf{1}$ & $%
		\mathbf{1}$ \\ \hline
		$U(1)_{Y}$ & $\frac{1}{6}$ & $\frac{2}{3}$ & $-\frac{1}{3}$ & $-\frac{1}{2}$
		& $-1$ & $0$ & $\frac{1}{2}$ & $\frac{1}{6}$ & $\frac{1}{6}$ & $\frac{2}{3}$
		& $\frac{2}{3}$ & $-\frac{1}{3}$ & $-\frac{1}{3}$ & $-\frac{1}{2}$ & $-\frac{%
			1}{2}$ & $-1$ & $-1$ & $0$ & $0$ & $0$ \\ \hline
		$U(1)'$ & $0$ & $0$ & $0$ & $0$ & $0$ & $0$ & $0$ & $q_{Q_{4}}$ & $%
		q_{Q_{4}} $ & $q_{u_{4}}$ & $q_{u_{4}}$ & $q_{d_{4}}$ & $q_{d_{4}}$ & $%
		q_{L_{4}}$ & $q_{L_{4}}$ & $q_{e_{4}}$ & $q_{e_{4}}$ & $q_{\nu _{4}}$ & $%
		q_{\nu _{4}}$ & $-q_{f_4}$ \\ \hline
	\end{tabular}
	\caption{Particle assigments under $SU(3)_{c}\times SU(2)_{L}\times U(1)_{Y}\times U(1)'$ gauge symmetry. $i=1,2,3$. The SM singlet scalars $\protect\phi_f $ ($f=Q,u,d,L,e$) have $U(1)'$ charges $-q_{f_4}=-q_{Q_{4,}u_{4},d_{4},L_{4},e_{4}}$.}
	\label{tab:model_content}
\end{table}

Although the $Z'$ couples only to the vector-like fourth family to start with, due to the mixing between SM fermions and those of the fourth vector-like family (arising from the Lagrangian below) the $Z'$ will get induced couplings to chiral
SM fermions.
After mixing, the model can allow for a viable dark matter candidate and operators crucial for explaining the $R_{K}$ and $R_{K^{\ast }} $ flavour anomalies\cite{Falkowski:2018dsl}. 
As we shall see, this setup can also generate CLFV signatures such as $\mu \rightarrow e\gamma $ and accommodate the experimental value of the anomalous muon and electron
magnetic dipole moments.

With the particle content, symmetries and charge assignments in Table \ref{tab:model_content}, the following renormalisable Lagrangian terms are available: 
\begin{align}
	\begin{split}
		\mathcal{L}_{Y} &=\sum_{i=1}^{3}\sum_{j=1}^{3}y_{ij}^{(u) }%
		\overline{Q}_{iL}\widetilde{H}u_{jR}+\sum_{i=1}^{3}\sum_{j=1}^{3}y_{ij}^{%
			(d) }\overline{Q}_{iL}Hd_{jR}+\sum_{i=1}^{3}%
		\sum_{j=1}^{3}y_{ij}^{(e)}\overline{L}_{iL}He_{jR}+%
		\sum_{i=1}^{3}\sum_{j=1}^{3}y_{ij}^{(\nu) }\overline{L}_{iL}%
		\widetilde{H}\nu _{jR} \\
		&+y_{4}^{(u)}\overline{Q}_{4L}\widetilde{H}u_{4R}+y_{4}^{		(d)}\overline{Q}_{4L}Hd_{4R}+y_{4}^{(e)}\overline{L}_{4L}HE_{4R}+y_{4}^{(\nu) }\overline{L}_{4L}\widetilde{H}\nu
		_{4R}\\
		&+\sum_{i=1}^{3}x_{i}^{(Q)}\phi _{Q}\overline{Q}_{Li}\tilde{Q}%
		_{4R}+\sum_{i=1}^{3}x_{i}^{(u)}\phi _{u}\overline{\tilde{u}}%
		_{4L}u_{Ri}+\sum_{i=1}^{3}x_{i}^{(d)}\phi _{d}\overline{\tilde{d%
		}}_{4L}d_{Ri}+\sum_{i=1}^{3}x_{i}^{(L)}\phi _{L}\overline{L}%
		_{Li}\tilde{L}_{4R}+\sum_{i=1}^{3}x_{i}^{(e)}\phi _{e}\overline{%
			\widetilde{E}}_{4L}e_{Ri} \\
		&+M_{4}^{Q}\overline{Q}_{4L}\tilde{Q}_{4R}+M_{4}^{u}\overline{\tilde{u}}_{4L}u_{4R}+M_{4}^{d}\overline{\tilde{d}}_{4L}d_{4R}+M_{4}^{L}\overline{L}_{4L}\tilde{L}%
		_{4R}+M_{4}^{E}\overline{\widetilde{E}}_{4L}E_{4R}+M_{4}^{\nu}\overline{\tilde{\nu}}_{4L}\nu _{4R}+H.c.
		\label{eqn:Yukawa_Lagrangian}
	\end{split}
\end{align}
where the requirement of $U(1)'$ invariance of the Yukawa interactions involving the fourth family yields the following constraints on the $U(1)'$ charges of fourth fermion families:
\begin{equation}
q_{Q_4}=q_{u_4}=q_{d_4}\hspace{1cm}q_{L_4}=q_{e_4}=q_{\nu_4}  
\end{equation} 
It is clear from Equation \eqref{eqn:Yukawa_Lagrangian} that fields in the 4th, vector-like family obtain masses from two sources; firstly, Yukawa terms involving the SM Higgs field such as $y_{4}^{(e)}\overline{L}_{4L}He_{4R}$ which get promoted to 
chirality flipping fourth family 
mass terms $M_4^C$ once the Higgs acquires a vev, and secondly from vector-like mass terms like $M_{4}^{L}\overline{L}_{4L}\tilde{L}_{4R}$ (these terms show up in lines 2 and 4 of Equation \eqref{eqn:Yukawa_Lagrangian} respectively). 
For the purposes of clarity, we shall treat $M_4^C$ and $M_{4}^{L}\overline{L}_{4L}\tilde{L}_{4R}$ as independent mass terms
in the analysis of the physical quantities of interest, rather than constructing the full fourth family mass matrix and 
diagonalising it, since such quantities rely on a chirality flip and are sensitive to $M_4^C$ rather than 
the vector-like masses $M_{4}^{L}\overline{L}_{4L}\tilde{L}_{4R}$. Spontaneous breaking of $U(1)'$ by the scalars $\phi_i$ spontaneously acquiring vevs gives rise to a massive $Z^{\prime }$ boson featuring couplings with the chiral and vector-like fermion fields. In the interaction basis such terms will be diagonal and of the following form:
\begin{align}
	\mathcal{L}_{Z^{\prime }}^{gauge}=g^{\prime }Z_{\mu }^{\prime }(
	\overline{Q}_{L}D_{Q}\gamma ^{\mu }{Q}_{L}+\overline{u}_{R}D_{u}\gamma ^{\mu
	}u_{R}+\overline{d}_{R}D_{d}\gamma ^{\mu }d_{R}+\overline{L}_{L}D_{L}\gamma
	^{\mu }L_{L}+\overline{e}_{R}D_{e}\gamma ^{\mu }e_{R} + \overline{\nu}_{R}D_{\nu}\gamma ^{\mu }\nu_{R})
	\label{eqn:Zprime_couplings_interaction_basis}
\end{align}
Here, $g'$ is the `pure' gauge coupling of $U(1)'$ and each of the $D$s are 4x4 matrices. However, only the fourth family has non-vanishing $U(1)'$ charges as per Table \ref{tab:model_content} and hence these matrices are given by:
\begin{equation}
	\begin{gathered}
		D_{Q}=\mathrm{diag}(0,0,0,q_{Q_4}),\quad
		D_{u}=\mathrm{diag}(0,0,0,q_{u_4}),\quad D_{d}=\mathrm{diag}(0,0,0,q_{d_4}),\\
		D_{L}=\mathrm{diag}(0,0,0,q_{L_4}),\quad D_{e}=\mathrm{diag}(0,0,0,q_{e_4}),\quad
		D_{\nu}=\mathrm{diag}(0,0,0,q_{\nu_4})
		\label{eqn:Zprime_coupling_matrices_interaction_basis}
	\end{gathered}
\end{equation}

At this stage, the SM quarks and leptons do not couple to the $Z'$. However, the Yukawa couplings detailed in Equation \eqref{eqn:Yukawa_Lagrangian} have no requirement to be diagonal. Before we can determine the full masses of the propagating vector-like states and SM fermions, we need to transform the field content of the model such that the Yukawa couplings become diagonal. Therefore, fermions in the mass basis (denoted by primed fields) are related to particles in the interaction basis by the following unitary transformations; 
\begin{equation}
	Q_{L}^{\prime }=V_{Q_{L}}{Q}_{L},\qquad u_{R}^{\prime }=V_{u_{R}}{u}%
	_{R},\qquad d_{R}^{\prime }=V_{d_{R}}{d}_{R},\qquad L_{L}^{\prime }=V_{L_{L}}%
	{L}_{L},\qquad e_{R}^{\prime }=V_{e_{R}}{e}_{R}
	,\qquad \nu_{R}^{\prime }=V_{\nu_{R}}{\nu}_{R}
	\label{eqn:field_transformations_to_mass_basis}
\end{equation}
This mixing induces couplings of SM mass eigenstate fermions to the massive $Z^{\prime }$ which can be expressed as follows
\begin{gather}
	\begin{split}
		D^{\prime }_Q=V_{Q_L}D_QV^{\dagger}_{Q_L}, \ \ \ \ 
		D^{\prime }_{u}=V_{u_R}D_uV^{\dagger}_{u_R}, \ \ \ \ 
		D^{\prime }_d=V_{d_R}D_dV^{\dagger}_{d_R},\\ 
		D^{\prime }_L=V_{L_L}D_LV^{\dagger}_{L_L}, \ \ \ \
		D^{\prime }_e=V_{e_R}D_eV^{\dagger}_{e_R}, \ \ \ \  
		D^{\prime }_\nu=V_{\nu_R}D_\nu V^{\dagger}_{\nu_R}
	\end{split}
	\label{eqn:coupling_conversion_to_mass_basis}
\end{gather}
Thus far all discussion of interactions and couplings has been general. In Sections \ref{sec:LFV_in_this_model} and \ref{sec:statistical_analysis}, we will prohibit mixing in some sectors to simplify our phenomenological analysis. In particular, we shall only consider induced $Z'$ couplings to the electron and muon.

\section{$Z'$ couplings to the electron and muon}
\label{sec:LFV_in_this_model}
In this paper we are particularly interested in the electron and muon g-2. We therefore take a minimal scenario and consider mixing only between first and second families of charged leptons, and ignore all quark and neutrino mixing, leading to a leptophillic $Z'$ model, in which the $Z'$ couples only to the electron, muon and their associated neutrinos. Therefore, only $V_{L_L}$ and $V_{e_R}$ will be non-diagonal, and LHC results will not constrain the $Z'$ mass as there is no direct coupling between SM quarks and the new vector boson, nor mixing between SM and vector-like quarks, because SM quarks are uncharged under $U(1)'$ as seen in Table \ref{tab:model_content}. Among the CLFV processes, we will focus on studying the $\mu\to e\gamma$ decay, which put tighter constrains than the $\tau\to\mu\gamma$ and $\tau\to e\gamma$ decays. For this reason, to simplify the parameter space, we also forbid the third family fermions from mixing with any other fermionic content. As such, all mixing at low energies can be expressed as per Equation \eqref{eqn:leptonic_mixing_matrices}.
\begin{gather}
	V_{L_L,e_R} =
	\begin{pmatrix}
	\cos\theta_{12}^{L,R} & \sin\theta_{12}^{L,R} & 0 & 0 \\ 
	-\sin\theta_{12}^{L,R} & \cos\theta_{12}^{L,R} & 0 & 0 \\ 
	0 & 0 & 1 & 0 \\ 
	0 & 0 & 0 & 1
	\end{pmatrix}
	\begin{pmatrix}
	\cos\theta_{14}^{L,R} & 0 & 0 & \sin\theta_{14}^{L,R} \\ 
	0 & 1 & 0 & 0 \\ 
	0 & 0 & 1 & 0 \\ 
	-\sin\theta_{14}^{L,R} & 0 & 0 & \cos\theta_{14}^{L,R}
	\end{pmatrix}
	\begin{pmatrix}
	1 & 0 & 0 & 0 \\ 
	0 & \cos\theta_{24}^{L,R} & 0 & \sin\theta_{24}^{L,R} \\ 
	0 & 0 & 1 & 0 \\ 
	0 & -\sin\theta_{24}^{L,R} & 0 & \cos\theta_{24}^{L,R}
	\end{pmatrix}
	\label{eqn:leptonic_mixing_matrices}
\end{gather}
The angles defined here take the theory from the interaction basis in Equation \eqref{eqn:Yukawa_Lagrangian} to the mass eigenbasis of primed fields introduced with Equation \eqref{eqn:field_transformations_to_mass_basis}. They directly parameterise the mixing between the 4th, vector-like family and the usual three chiral families of SM fermions. Such mixing parameters will cause the $D'$ matrices from Equation \eqref{eqn:coupling_conversion_to_mass_basis} to become off-diagonal. This incites couplings between the massive $Z'$ vector boson and the SM leptons, suppressed by the mixing angles. These mixing angles can be expressed in terms of parameters from the Lagrangian (Equation \eqref{eqn:Yukawa_Lagrangian}), as per Equation \eqref{eqn:angles_from_lag_params} \cite{King:2017anf}.
\begin{gather}
	\tan\theta_{14}^L = \frac{x_1^{(L)}\langle\phi_L\rangle}{M_4^L}\,\,,\qquad \tan\theta_{24}^L = \frac{x_2^{(L)}\langle\phi_L\rangle}{\sqrt{\big(x_1^{(L)}\langle\phi_L\rangle\big)^2+\big(M_4^L\big)^2}}
	\label{eqn:angles_from_lag_params}
\end{gather}
With the restrictions defined in Equation \eqref{eqn:leptonic_mixing_matrices} and above, all of the relevant couplings between the massive $Z^{\prime}$ and fermions in the mass basis of propagating fields can be determined as the following:
\begin{equation}
\mathcal{L}_{Z^{\prime }}^{gauge}=Z_{\mu }^{\prime } \overline{l}_{L,R}(g_{L,R})_{ll'}\gamma ^{\mu }l'_{L,R}
\end{equation}
where $l,l'=e, \mu, E$, the mass eigenstate leptons electron, muon and vector-like lepton respectively with the following couplings to the massive $Z'$ boson:
\begin{align}
	(g_{L,R})_{\mu\mu} &= g'q_{L_4,e_4}\Big(\cos\theta_{12}^{L,R}\sin\theta_{24}^{L,R}-\cos\theta_{24}^{L,R}\sin\theta_{12}^{L,R}\sin\theta_{14}^{L,R}\Big)^2\label{eqn:mu_mu_zprime_coupling}\\[7pt]
	(g_{L,R})_{ee} &= g'q_{L_4,e_4}\Big(\sin\theta_{12}^{L,R}\sin\theta_{24}^{L,R}+\cos\theta_{12}^{L,R}\cos\theta_{24}^{L,R}\sin\theta_{14}^{L,R}\Big)^2\label{eqn:e_e_zprime_coupling}\\[7pt]
	(g_{L,R})_{EE} &= g'q_{L_4,e_4}\Big(\cos\theta_{14}^{L,R}\Big)^2\Big(\cos\theta_{24}^{L,R}\Big)^2\label{eqn:E_E_zprime_coupling}\\[7pt]
	(g_{L,R})_{eE} &= g'q_{L_4,e_4}\cos\theta_{14}^{L,R}\cos\theta^{L,R}_{24}\Big(\sin\theta_{12}^{L,R}\sin\theta_{24}^{L,R}+\cos\theta_{12}^{L,R}\cos\theta_{24}^{L,R}\sin\theta_{14}^{L,R}\Big)\label{eqn:e_E_zprime_coupling}\\[7pt]
	(g_{L,R})_{\mu E} &= g'q_{L_4,e_4}\cos\theta_{14}^{L,R}\cos\theta_{24}^{L,R}\Big(\cos\theta_{12}^{L,R}\sin\theta_{24}^{L,R}-\cos\theta_{24}^{L,R}\sin\theta_{12}^{L,R}\sin\theta_{14}^{L,R}\Big)\label{eqn:mu_E_zprime_coupling}\\[7pt]
	(g_{L,R})_{\mu e} &= g'q_{L_4,e_4}\Big(\sin\theta_{12}^{L,R}\sin\theta_{24}^{L,R}+\cos\theta_{12}^{L,R}\cos\theta_{24}^{L,R}\sin\theta^{L,R}_{14}\Big)\Big(\cos\theta_{12}^{L,R}\sin\theta_{24}^{L,R}-\cos\theta_{24}^{L,R}\sin\theta_{12}^{L,R}\sin\theta_{14}^{L,R}\Big)
	\label{eqn:mu_e_zprime_coupling}
\end{align}
It is important to note that only the first and second family of SM leptons $e,\mu$ couple to the massive $Z^{\prime}$, 
with their non-universal and flavour changing couplings controlled by the mixing angles $\theta^{L,R}_{14}, \theta^{L,R}_{24}$
with the vector-like family. Throughout the remainder of this work, we assume that $g'q_{L4,e4}=1$ for simplicity.

\subsection{Muon decay to electron plus photon}
\begin{figure}[tbp]
	\centering
	\includegraphics[width=0.7\textwidth]{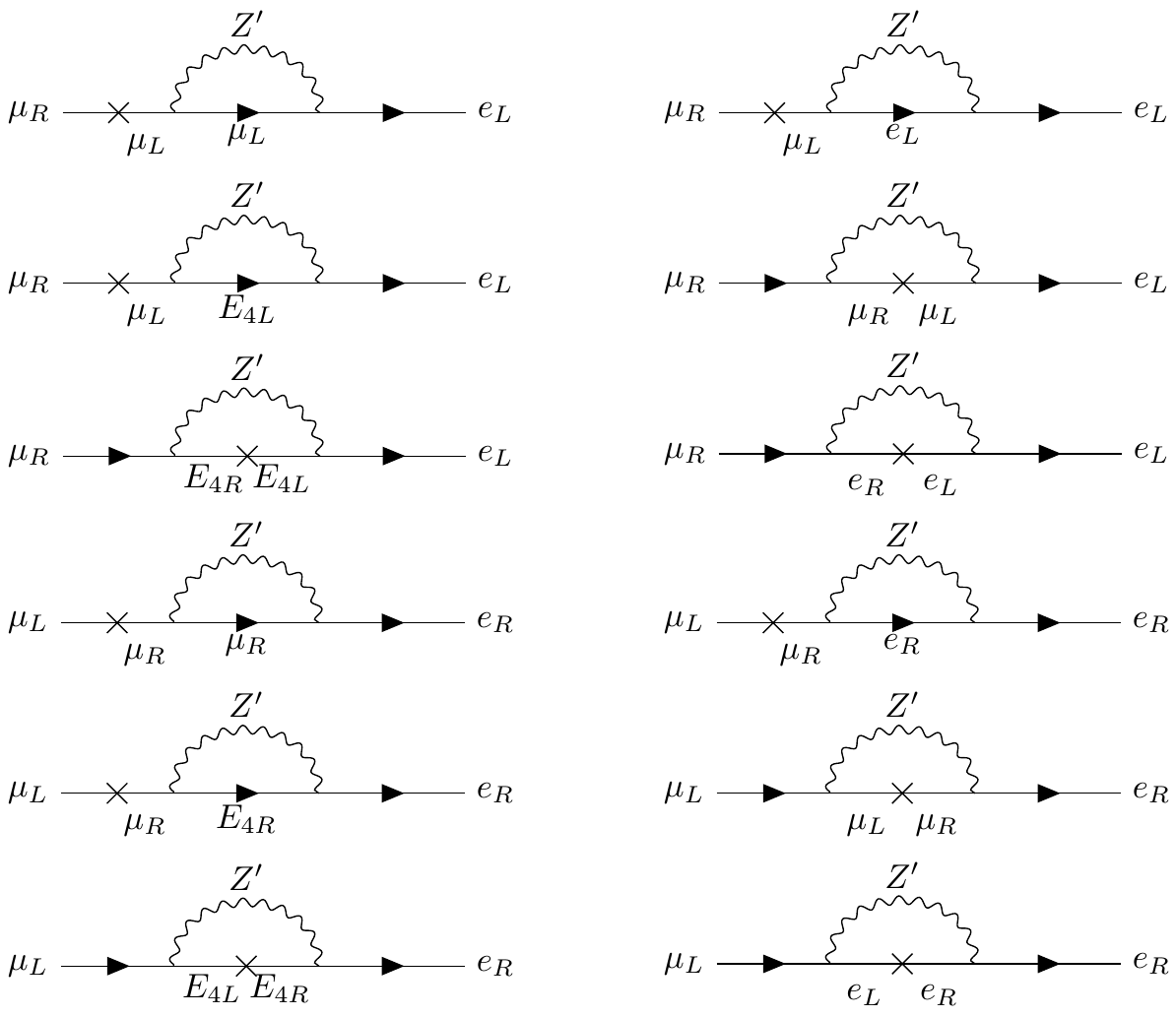}
	\caption{Feynamn diagrams contributing to the $\mu \rightarrow e\gamma $ decay. Note that 
	these diagrams all rely on a chirality flipping mass (LR). Where the chirality flip involves the fourth family, 
	the relevant mass is $M_4^C$.}
	\label{fig:muegamma_feynman_diagrams}
\end{figure}

In this subsection we study charged lepton flavor violating process $\mu\to e\gamma$ in the context of our BSM scenario. It is worth mentioning that a future observation of the $\mu\to e\gamma$ decay will be indisputable evidence of physics beyond the SM . The SM does predict non-zero branching ratios for the processes $\mu\to e\gamma$, $\tau\to\mu\gamma$ and $\tau\to e\gamma$, but such predictions are several orders of magnitude below projected experimental sensitivities \cite{Lindner:2016bgg,Calibbi:2017uvl}. The $\mu\rightarrow e\gamma$ decay rate is enhanced with respect to the SM by additional contributions due to virtual $Z^\prime$ and charged exotic lepton exchange at the one-loop level. General $l_{i}\rightarrow l_{j}\gamma$ decay can be described by the following effective operator \cite{Lindner:2016bgg}: 
\begin{equation}
	\mathcal{L}_{EFT}=\frac{\mu _{ij}^{M}}{2}\overline{l}_{i}\sigma ^{\mu \nu}l_{j}F_{\mu \nu }+\frac{\mu_{ij}^{E}}{2}i\overline{l}_{i}\gamma ^{5}\sigma^{\mu \nu }l_{j}F_{\mu \nu }
	\label{eqn:anomalous_moments_effective_lagrangian}
\end{equation}
where $F_{\mu\nu}$ denotes the electromagnetic field strength tensor, $\mu _{ij}^{E}$ and $\mu _{ij}^{M}$ are the transition electric and magnetic moments, respectively and $i,j=1,2,3$ denote family indices. Diagonal elements in the transition magnetic moment $\mu _{ij}^{M}$ give rise to the anomalous dipole moments $\Delta a_{l}=\frac{1}{2}(g_{l}-2)$ of leptons, whilst off-diagonal elements in the transition moments contribute to the $l_{i}\rightarrow l_{j}\gamma $ decay amplitude. Based on the effective Lagrangian in Equation \eqref{eqn:anomalous_moments_effective_lagrangian}, one has that the amplitude for a generic lepton decay $f_{1}\rightarrow f_{2}\gamma $ has the form \cite{Lavoura:2003xp}: 
\begin{equation}
	\mathcal{A} =\, e\varepsilon _{\mu }^{\ast }(q)\overline{v }_{2}(p_{2})\left[i\sigma^{\mu\nu}q_{\nu}(\sigma_{L}P_{L}+\sigma_{R}P_{R})\right]u_{1}(p_{1})	\label{eqn:LFV_decay_amplitude}
\end{equation}
where $\sigma _{L}$ and $\sigma _{R}$ are numerical quantities with dimension of inverse mass that can be expressed in terms of loop integrals \cite{Lavoura:2003xp}. $u_1$ and $v_2$ are spinors, furthermore, we have the following relations: 
\begin{equation}
	\begin{gathered}
		\sigma ^{\mu \nu } =\frac{i}{2}\left[\gamma^{\mu},\gamma^{\nu}\right],\hspace{1.5cm}P_{L,R}=\frac{1}{2}(1\mp\gamma_{5}),\hspace{ 1.5cm}q=p_{1}-p_{2} 
	\end{gathered}  
	\label{eqn:operators_and_projectors}
\end{equation}
In such a general case, the decay rate expression for the $\mu\to e\gamma$ process is the following \cite{Lavoura:2003xp,Chiang:2011cv,Lindner:2016bgg,Raby:2017igl}:
\begin{equation}
	\Gamma(\mu\rightarrow e\gamma)=\frac{\alpha_{em}}{1024\pi^{4}}\frac{m_{\mu }^{5}}{M_{Z^{\prime }}^{4}}(\left\vert \widetilde{\sigma }_{L}\right\vert ^{2}+\left\vert \widetilde{\sigma }_{R}\right\vert^{2})
	\label{eqn:mu_e_gamma_decay_rate_prediction}
\end{equation}
where $\widetilde{\sigma }_{L}$ and $\widetilde{\sigma }_{R}$ are given by: 
\begin{align}
	\begin{split}
	\widetilde{\sigma }_{L}& =\sum_{a=e,\mu,E}\left[(g_{L})_{ea}(g_{L})_{a\mu}F(x_{a})+\frac{m_{a}}{m_{\mu }}(g_{L})_{ea}(g_{R})_{a\mu}G(x_{a})\right] , \\
	\widetilde{\sigma }_{R}& =\sum_{a=e,\mu,E}\left[ (g_{R})_{ea}(g_{R})_{a\mu}F(x_{a})+\frac{m_{a}}{m_{\mu }}(g_{R})_{ea}(g_{L})_{a\mu}G(x_{a})\right],\hspace{1.5cm}x_{a}=\frac{m_{a}^{2}}{M_{Z^{\prime }}^{2}}
	\label{eqn:contributions_to_muegamma_(sigmas)}
	\end{split}
\end{align}
$F(x)$ and $G(x)$ are loop functions related to the Feynman diagrams for $\mu\rightarrow e\gamma$ as per Figure \ref{fig:muegamma_feynman_diagrams}, and have the functional form given in Equation \eqref{eqn:loop_functions}. $g_{L,R}$ are couplings in the fermion mass basis, as detailed in Equations \eqref{eqn:mu_mu_zprime_coupling} through \eqref{eqn:mu_e_zprime_coupling}. $m_a$ here corresponds to the full propagating mass of the vector-like partners. In the approximation where the vector like mass $M_4^L$ is always much greater than the chirality-flipping mass $M_4^C$ ($M_4^L\gg M_4^C$) that we will adopt here, this full propagating mass is almost equivalent to the vector-like mass. Therefore when $a=E$, we approximate $m_E\simeq M_4^L$. The loop functions are given by \cite{Raby:2017igl}: 
\begin{equation}
	\begin{gathered} 
		F(x)=\frac{5x^{4}-14x^{3}+39x^{2}-38x-18x^{2}\ln x+8}{ 12(1-x)^{4}}, \\ G(x)=\frac{x^{3}+3x-6x\ln x-4}{2(1-x)^{3}} 
	\end{gathered}  
\label{eqn:loop_functions}
\end{equation}
Equation \eqref{eqn:mu_e_gamma_decay_rate_prediction} has some generic features; the loop function $F(x)$ varies between 0.51 and 0.67 when $x$ is varied in the range $10^{-3}\le x \le 2$, whilst in the same region, $G(x)$ varies between -1.98 and -0.84. Consequently, in the case of charged fermions running in loops, contributions proportional to $G(x)$ will likely dominate over those proportional to $F(x)$. The dominant contributions involve left-right and right-left $Z^{\prime }$ couplings, whereas the subleading ones include either left-left or right-right couplings. Dividing Equation \eqref{eqn:mu_e_gamma_decay_rate_prediction} by the known decay rate of the muon yields a prediction for the $\mu\rightarrow e\gamma$ branching fraction \cite{Lavoura:2003xp,Chiang:2011cv,Lindner:2016bgg,Raby:2017igl}: 
\begingroup
\addtolength{\jot}{5pt}
\begin{align}
	\begin{split}
		\operatorname{BR}(\mu \rightarrow e \gamma) = \frac{\alpha}{1024\pi^4}&\frac{m_{\mu}^5}{M_{Z^{\prime}}^4 \Gamma_{\mu}} \Bigg[ \Big\vert(g_L)_{\mu\mu}(g_L)_{\mu e} F(x_\mu) + (g_L)_{\mu E} (g_L)_{eE} F(x_{E})+ (g_L)_{\mu e} (g_L)_{e e} F(x_e) \\
		&+ \frac{m_{\mu}}{m_{\mu}}(g_L)_{\mu e} (g_R)_{\mu \mu} G(x_\mu) + \frac{M_4^C}{m_{\mu}}(g_L)_{eE} (g_R)_{\mu E} G(x_{E}) + \frac{m_e}{m_{\mu}}(g_L)_{e e} (g_R)_{\mu e} G(x_e) \Big\vert^2 \\
		&+ \Big\vert (g_R)_{\mu \mu} (g_R)_{\mu e} F(x_\mu) + (g_R)_{\mu E} (g_R)_{eE} F(x_{E}) + (g_R)_{\mu e} (g_R)_{e e} F(x_e)\\
		&+\frac{m_{\mu}}{m_{\mu}}(g_R)_{\mu e} (g_L)_{\mu \mu} G(x_\mu)+ \frac{M_4^C}{m_{\mu}}(g_R)_{eE} (g_L)_{\mu E} G(x_{E}) + \frac{m_e}{m_{\mu}}(g_R)_{e e} (g_L)_{\mu e} G(x_e) \Big\vert^2 \Bigg]
		\label{eqn:mu_e_gamma}
	\end{split}
\end{align}
\endgroup

where the total muon decay width is $\Gamma _{\mu }=\frac{G_{F}^{2}m_{\mu}^{5}}{192\pi ^{3}}=3\times 10^{-19}\text{GeV}$. The mass $M_4^C$ that appears in the Feynman diagrams  with a chirality flip on the 4th family fermions $E_4$ (Figure \ref{fig:muegamma_feynman_diagrams}, 5th and 11th diagrams) is \emph{not} the vector-like mass, but instead arises from the Yukawa-like couplings from Equation \eqref{eqn:Yukawa_Lagrangian}, $M_4^C=y^{(e)}_{44}v_{\phi}$, where $v_{\phi}$ is the vacuum expectation value of the SM Higgs field, which acquires a vev and spontaneously breaks electroweak symmetry in the established manner. Under the assumption that $M_4^C>m_\mu$, such terms proportional to the chirality flipping mass in Equation \eqref{eqn:mu_e_gamma} give by far the largest contributions to $\mu\rightarrow e\gamma$. The experimental limit on $\operatorname{BR}(\mu\rightarrow e\gamma)$ is determined from non-observation at the MEG experiment at a 90\% confidence level\cite{TheMEG:2016wtm,Tanabashi:2018oca}:
\begin{equation}
\operatorname{BR}(\mu\rightarrow e\gamma) < 4.2\times10^{-13}
\label{eqn:muegamma_limit}
\end{equation}

\subsection{Anomalous magnetic moment of the muon $\Delta a_\mu$}
\begin{figure}[tbp]
	\centering
	\includegraphics[width=0.75\textwidth]{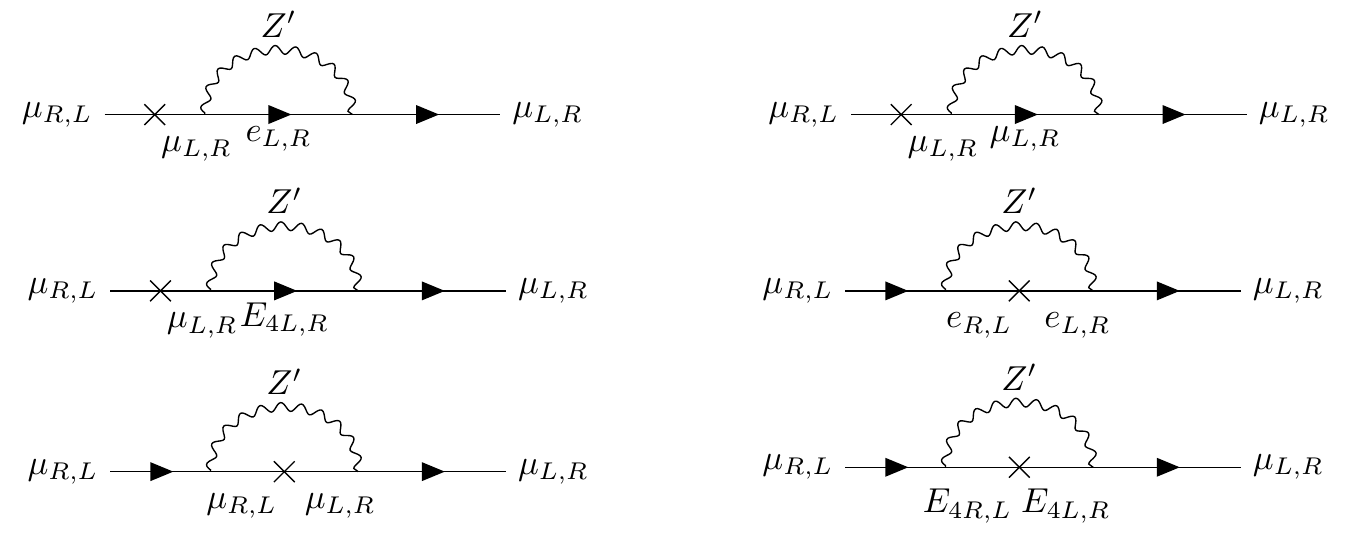}
	\caption{Feyman diagrams contributing to the muon $(g-2)_{\mu}$}
	\label{fig:muong2_feynman_diagrams}
\end{figure}

In this subsection we study the muon anomalous magnetic moment in the context of our BSM scenario. In a model such as this, the Feynman diagrams for $\mu\rightarrow e\gamma$ are easily modified to give contributions to the anomalous magnetic moment of the muon as per Figure \ref{fig:muong2_feynman_diagrams}. The prediction for such an observable in our model therefore takes the form \cite{Raby:2017igl}:
\begingroup
\addtolength{\jot}{5pt}
\begin{align}
	\begin{split}
		\Delta a_{\mu}^{Z^{\prime}} = -\frac{m_{\mu}^2}{8\pi^2 M_{Z^{\prime}}^2} \bigg[ \big( \vert &(g_L)_{\mu\mu} \vert^2 + \vert (g_R)_{\mu\mu} \vert^2 \big) F(x_{\mu}) 
		+ \big( \vert (g_L)_{\mu E} \vert^2 + \vert (g_R)_{\mu E} \vert^2 \big) F(x_{E}) \\
		&+ \big( \vert (g_L)_{\mu e} \vert^2 + \vert (g_R)_{\mu e} \vert^2 \big) F(x_{e}) 
		+ \operatorname{Re}\big( (g_L)_{\mu \mu} (g_R^*)_{\mu \mu} \big) G(x_{\mu}) \\
		&+ \operatorname{Re}\big( (g_L)_{\mu E} (g_R^*)_{\mu E} \big) \frac{M_4^C}{m_{\mu}} G(x_{E}) 
		+ \operatorname{Re}\big( (g_L)_{\mu e} (g_R^*)_{\mu e} \big) \frac{m_{e}}{m_{\mu}} G(x_{e}) \bigg]
	\label{eqn:muon_g-2_contributions}
	\end{split}
\end{align}
\endgroup

Once more, the dominant terms will be those proportional to the enhancement factor of $\frac{M_4^C}{m_\mu}$, corresponding to the final diagram in Figure \ref{fig:muong2_feynman_diagrams}, provided $M_4^C>m_\mu$. Recent experimental evidence has shown that the muon magnetic moment as measured by the E821 experiment at BNL is at around a 3.5$\sigma$ deviation from the SM prediction\cite{Hagiwara:2011af,Nomura:2018lsx,Nomura:2018vfz,Bennett:2006fi,Davier:2010nc,Davier:2017zfy,Davier:2019can}: 
\begin{equation}
	(\Delta a_{\mu })_{\exp}=(26.1\pm 8)\times10^{-10}
	\label{eqn:experimental_muon_g-2_data}
\end{equation}

\subsection{Anomalous magnetic moment of the electron $\Delta a_{e}$}
\begin{figure}[tbp]
	\centering
	\includegraphics[width=0.75\textwidth]{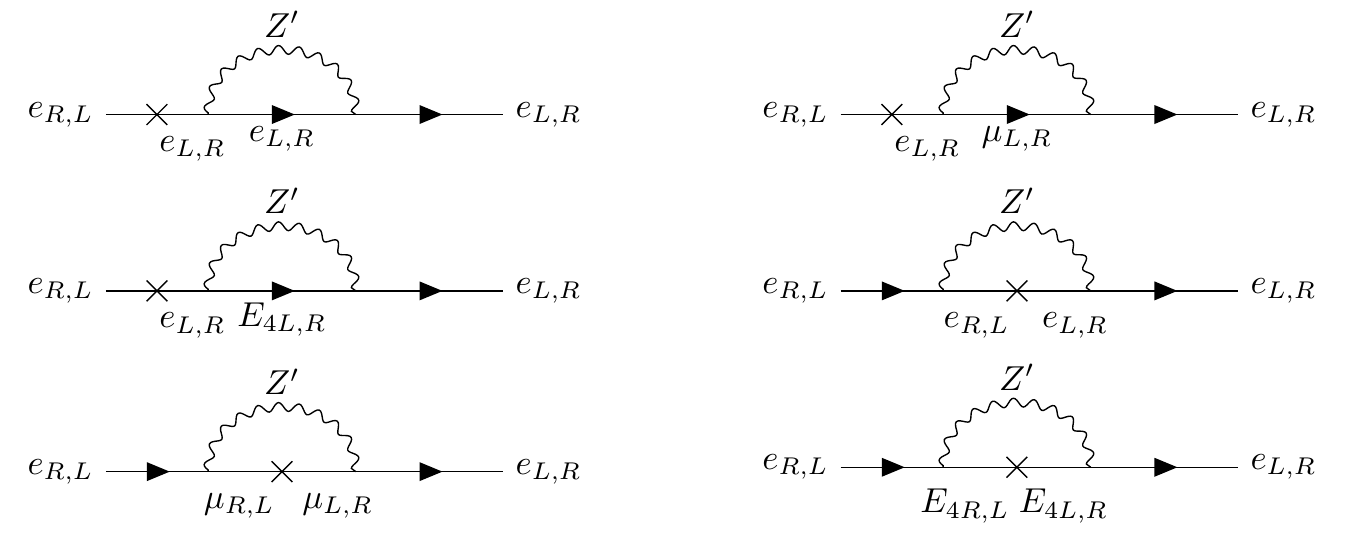}
	\caption{Feynamn diagrams contributing to the electron $g-2$}
	\label{fig:electrong-2_feynman_diagrams}
\end{figure}

Analogously to the muon, there is also an amendment to the electron $(g-2)_e$ in this scenario, from Feynman diagrams given in Figure \ref{fig:electrong-2_feynman_diagrams}. The analytic expression for $\Delta a_e$ is the following \cite{Raby:2017igl}:
\begingroup
\addtolength{\jot}{5pt}
\begin{align}
	\begin{split}
		\Delta a_{e}^{Z^{\prime}} = -\frac{m_{e}^2}{8\pi^2 M_{Z^{\prime}}^2} \bigg[ \big( \vert &(g_L)_{ee} \vert^2 + \vert (g_R)_{ee} \vert^2 ) F(x_{e}) 
		+ \big( \vert (g_L)_{e \mu} \vert^2 + \vert (g_R)_{e \mu} \vert^2 \big) F(x_{\mu}) \\
		&+ \big( \vert (g_L)_{eE} \vert^2 + \vert (g_R)_{eE} \vert^2 \big) F(x_{E}) 
		+ \operatorname{Re}\big( (g_L)_{ee} (g_R^*)_{ee} \big) \frac{m_{e}}{m_{e}} G(x_{e}) \\
		&+ \operatorname{Re}\big( (g_L)_{e \mu} (g_R^*)_{e \mu} \big) \frac{m_{\mu}}{m_{e}} G(x_{\mu}) 
		+ \operatorname{Re}\big( (g_L)_{eE} (g_R^*)_{eE} \big) \frac{M_4^C}{m_{e}} G(x_{E}) \bigg]
	\end{split}
	\label{eqn:electron_g-2_contributions}
\end{align}
\endgroup

As per the muon moment, if $M_4^C>m_\mu$ the largest contribution to the electron moment will be the final term in Equation \eqref{eqn:electron_g-2_contributions}, corresponding to the last diagram in Figure \ref{fig:electrong-2_feynman_diagrams}. The most recent experimental result of the $(g-2)_e$, obtained from measurement of the fine structure constant of QED, shows a $2.5\sigma$ deviation from the SM, similarly to the muon magnetic moment\cite{Parker:2018vye}: 
\begin{equation}
	(\Delta a_e)_{\text{exp}} = (-0.88\pm0.36)\times 10^{-12}
	\label{eqn:experimental_electron_g-2_data}
\end{equation}
Notice especially that Equations \eqref{eqn:experimental_muon_g-2_data} and \eqref{eqn:experimental_electron_g-2_data} have deviations from the SM in opposite directions, therefore explaining both phenomena simultaneously can be difficult for a given model to achieve.

\subsection{Neutrino trident production}
So-called trident production of neutrinos by process $\nu_{\mu}\gamma^{\ast}\rightarrow\nu_{\mu}\mu^{+}\mu^{-}$ through nuclear scattering is also relevant. The Feynamn diagram contributing to neutrino trident production in our model is shown in Figure \ref{fig:trident_feynman_diagrams}. This process constrains the following effective four lepton interaction, which in this scenario arises from leptonic $Z^{\prime}$ interactions \cite{Geiregat:1990gz,Mishra:1991bv,Altmannshofer:2014pba}: 
\begin{equation}
	\Delta\mathcal{L}_{eff}\supset-\frac{(g_L)_{\mu\mu}^{2}}{2M_{Z^{\prime}}^{2}}(\overline{\mu}_{L}\gamma^{\lambda}\mu_{L})(\overline{\nu}_{\mu L}\,\gamma_{\lambda}\,\nu_{\mu L}) -\frac{(g_R)_{\mu\mu}(g_L)_{\mu\mu}}{2M_{Z^{\prime}}^{2}}(\overline{\mu}_{R}\gamma^{\lambda}\mu_{R})(\overline{\nu}_{\mu L}\,\gamma_{\lambda}\,\nu_{\mu L})
	\label{eqn:trident_effective_Lagrangian}
\end{equation}
Said coupling is constrained as in the $SU(2)_L$ symmetric SM, left-handed muons and left-handed muon neutrinos couple identically to the $Z'$ vector boson. Experimental data on neutrino trident production $\nu_{\mu}\gamma^{\ast}\rightarrow\nu_{\mu}\mu^{+}\mu^{-}$ yields the following constraint at $95\%$ CL \cite{Falkowski:2017pss}: 
\begin{figure}[tbp]
	\centering
	\includegraphics[width=0.75\textwidth]{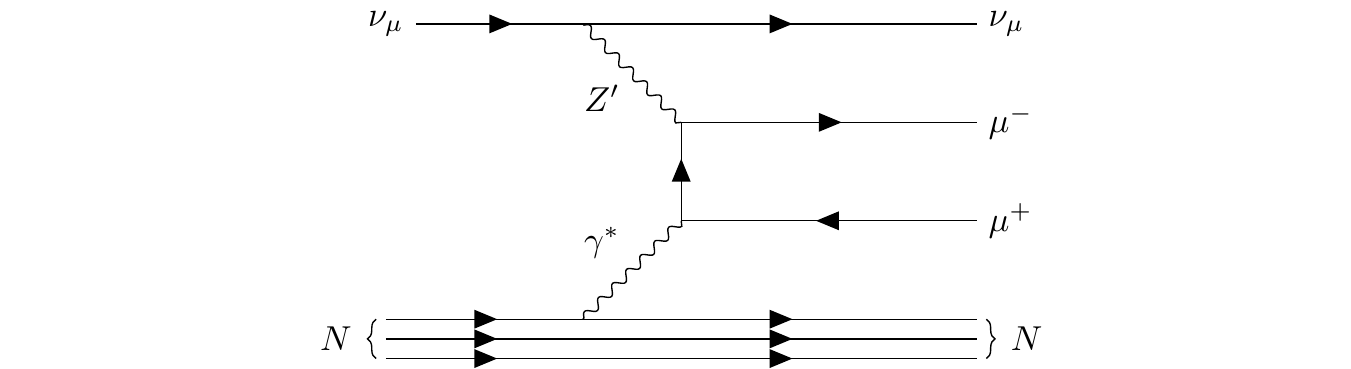}
	\caption{Feynamn diagram contributing to neutrino trident production, $N$ denotes a nucleus.}
	\label{fig:trident_feynman_diagrams}
\end{figure}
\begin{equation}
	-\frac{1}{(390\text{GeV})^{2}}\,\lesssim\frac{(g_L)_{\mu\mu}^{2}+(g_L)_{\mu\mu}(g_R)_{\mu\mu}}{M_{Z^{\prime}}^{2}}\lesssim\frac{1}{(370\text{GeV})^{2}}
	\label{eqn:trident_constraint_on_effective_coupling}
\end{equation}
This limit can be applied to the model's parameter space in a similar manner to other CLFV constraints discussed previously.

\section{Analytic arguments for $(g-2)_\mu$, $(g-2)_e$ and $\operatorname{BR}(\mu\rightarrow e\gamma)$}
\label{sec:analytics}
In order to gain an analytic understanding of the interplay between $(g-2)_\mu$, $(g-2)_e$ and $\operatorname{BR}(\mu\rightarrow e\gamma)$, in this section we shall make some simplifying assumptions about the parameters appearing in Equations \eqref{eqn:muon_g-2_contributions}, \eqref{eqn:electron_g-2_contributions} and \eqref{eqn:mu_e_gamma}.
If we assume large fourth family chirality flipping masses
$M_4^C\gg m_\mu$, then the expressions for these phenomena reduce to a minimal number of terms, all proportional to $M_4^C$. Furthermore, we assume that left- and right- handed couplings are related by some real, positive constants $k_1$ and $k_2$ defined thus:
\begin{align}
\begin{split}
(g_L)_{\mu E} = g_{\mu E}, \quad&\quad (g_R)_{\mu E} = k_1g_{\mu E},\\
(g_L)_{eE} = g_{eE}, \quad&\quad (g_R)_{eE} = -k_2g_{eE}
\label{eqn:analytic_coupling_assumptions}
\end{split}
\end{align}
The final coupling in Equation \eqref{eqn:analytic_coupling_assumptions} is defined with a sign convention such that, seeing as it is known numerically that the $G$ loop function is always negative, we automatically recover the correct signs for all of our observables. We also define the following prefactor constants to further simplify our expressions:
\begin{gather}
C_1 = \frac{\alpha}{1024\pi^2}\frac{m_\mu^5}{M_{Z'}^4\Gamma_\mu}\; ,\quad C_2 = \frac{m_\mu^2}{8\pi^2M_{Z'}^2}\; ,\quad C_3 = \frac{m_e^2}{8\pi^2M_{Z'}^2}
\label{eqn:simplified_prefactor_constants}
\end{gather}
Under such assumptions, Equations \eqref{eqn:muon_g-2_contributions}, \eqref{eqn:electron_g-2_contributions} and \eqref{eqn:mu_e_gamma} reduce to the following:
\begin{gather}
\operatorname{BR}(\mu\rightarrow e\gamma) = C_1\Bigg(\Big\lvert\frac{M_4^C}{m_\mu}k_1g_{eE}g_{\mu E}G(x_E)\Big\rvert^2+\Big\lvert\frac{M_4^C}{m_\mu}k_2g_{eE}g_{\mu E}G(x_E)\Big\rvert^2\Bigg)
\label{eqn:simplified_mu_e_gamma_expression}
\\[2ex]
\operatorname{|\Delta a_\mu|} = C_2k_1g_{\mu E}^2\frac{M_4^C}{m_\mu}|G(x_E)|
\label{eqn:simplified_muong2_expression}
\\[2ex]
\operatorname{|\Delta a_e|} = C_3k_2g_{eE}^2\frac{M_4^C}{m_e}|G(x_E)|
\label{eqn:simplified_electrong2_expression}
\end{gather}
We can then invert Equations \eqref{eqn:simplified_muong2_expression} and \eqref{eqn:simplified_electrong2_expression} to obtain expressions for the couplings in terms of the observables as per Equation \eqref{eqn:couplings_from_observables}.
\begin{gather}
g_{\mu E} = \sqrt{\frac{|\Delta a_\mu|}{C_2k_1}\frac{1}{|G(x_E)|}\frac{m_\mu}{M_4^C}}\;,\quad\quad g_{eE} = \sqrt{\frac{|\Delta a_e|}{C_3k_2}\frac{1}{|G(x_E)|}\frac{m_e}{M_4^C}}
\label{eqn:couplings_from_observables}
\end{gather}
Substituting into the flavour violating muon decay in Equation \eqref{eqn:simplified_mu_e_gamma_expression} and expanding the constants defined earlier yields:
\begin{align}
	\operatorname{BR}(\mu\rightarrow e\gamma) = \frac{\alpha\pi^2}{16}\frac{(k_1^2+k_2^2)}{k_1k_2}|\Delta a_\mu||\Delta a_e|\frac{m_\mu^2}{\Gamma_\mu m_e}
	\label{eqn:really_simplified_mu_e_gamma_expression}
\end{align}
independently of $M_{Z'}$ and $M_4^C$ which cancel.
Rearranging Equation \ref{eqn:really_simplified_mu_e_gamma_expression} and setting the physical quantities 
$|\Delta a_\mu|$, $|\Delta a_e|$ equal to their desired 
central values, yields a simple condition on $r=k_1/k_2$ in order to satisfy the bound on 
$\operatorname{BR}(\mu\rightarrow e\gamma)$:
\begin{align}
\| r+\frac{1}{r} \|< 5.57\times10^{-10}
\end{align}
Since the left hand side is minimised for $r=1$, 
the bound on $\operatorname{BR}(\mu\rightarrow e\gamma)$
can never be satisfied while accounting for $(g-2)_\mu$, $(g-2)_e$
(although clearly it is possible to satisfy it with either $(g-2)_\mu$ or $(g-2)_e$ but not both).
However this conclusion is based on the assumption that the physical quantities are dominated by the diagrams involving the 
chirality flipping fourth family masses $M_4^C\gg m_\mu$. In order to relax this assumption, a more complete 
analysis of the parameter space is required, one that considers all relevant terms in our expressions for observables in a numerical exploration of the parameter space. Such investigations are detailed in Section \ref{sec:statistical_analysis}.

\section{Numerical Analysis of the Fermiophobic $Z'$ Model}
\label{sec:statistical_analysis}
Given the expressions for observables that we have outlined above, we use these phenomena to constrain the parameter space of the model. As mentioned, a minimal parameter space is considered here, limiting mixing to the lepton sector and omitting the third chiral family from any mixing. From coupling expressions in Section \ref{sec:LFV_in_this_model}, the angular mixing parameters such as $\theta_{24L}$ and particle masses form a minimal parameter space for this model. We set direct mixing between the electron and muon ($\theta_{12L,R}$) to be vanishing for all tests, as even small direct mixing can easily violate the strict MEG constraint on $\operatorname{BR}(\mu\rightarrow e\gamma)$.

\subsection{Anomalous muon magnetic moment}
\begin{figure}[H]
	\centering
	\begin{subfigure}{0.48\textwidth}
		\includegraphics[width=1.0\textwidth]{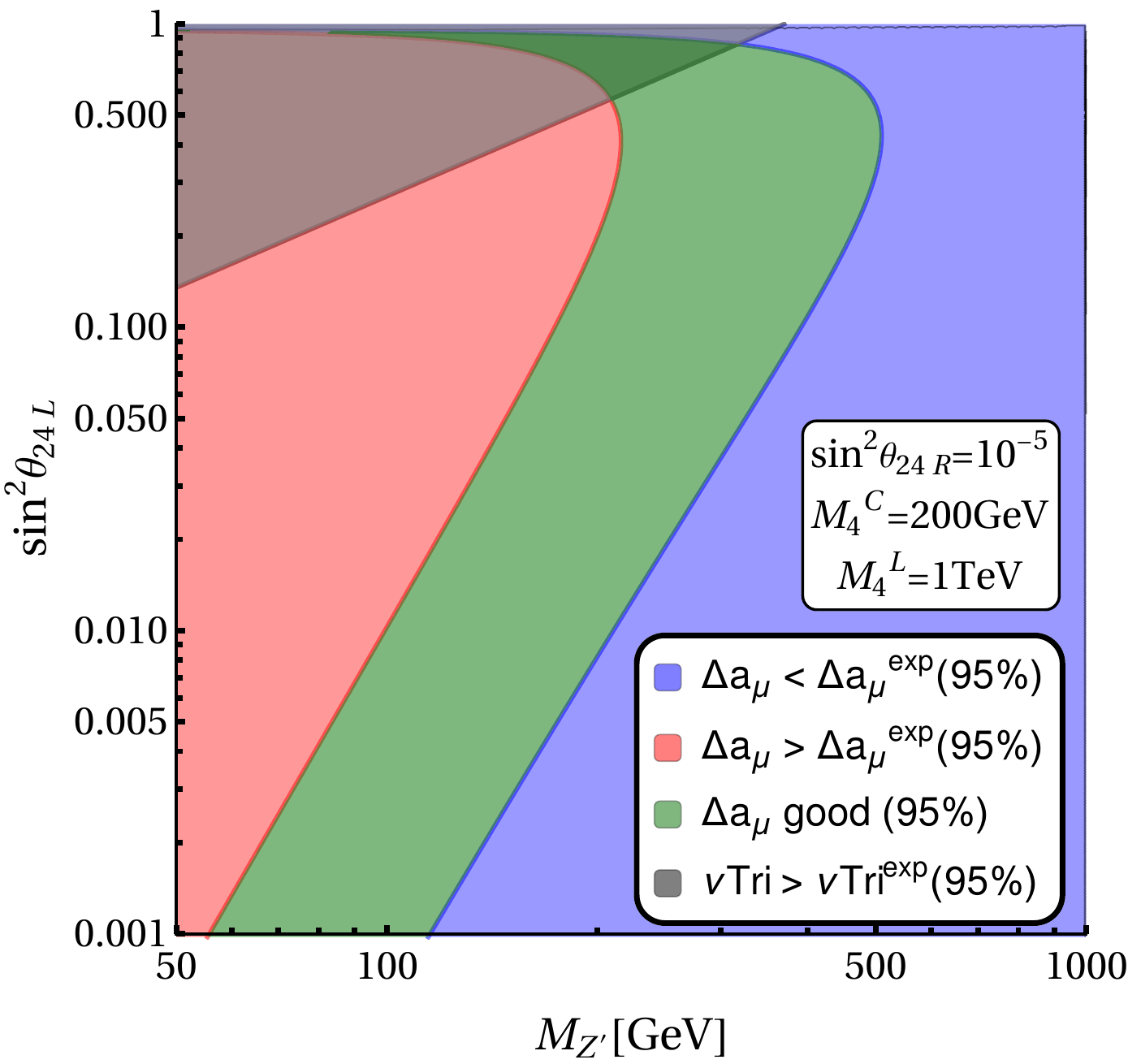}
		\captionsetup{width=0.8\linewidth}
		\caption{Trident exclusion and regions of $\Delta a_\mu$, with a fixed $\sin^2\theta_{24R}$.}
		\label{fig:muong-2_and_trident1}
	\end{subfigure}
	\begin{subfigure}{0.48\textwidth}
		\includegraphics[width=0.96\textwidth]{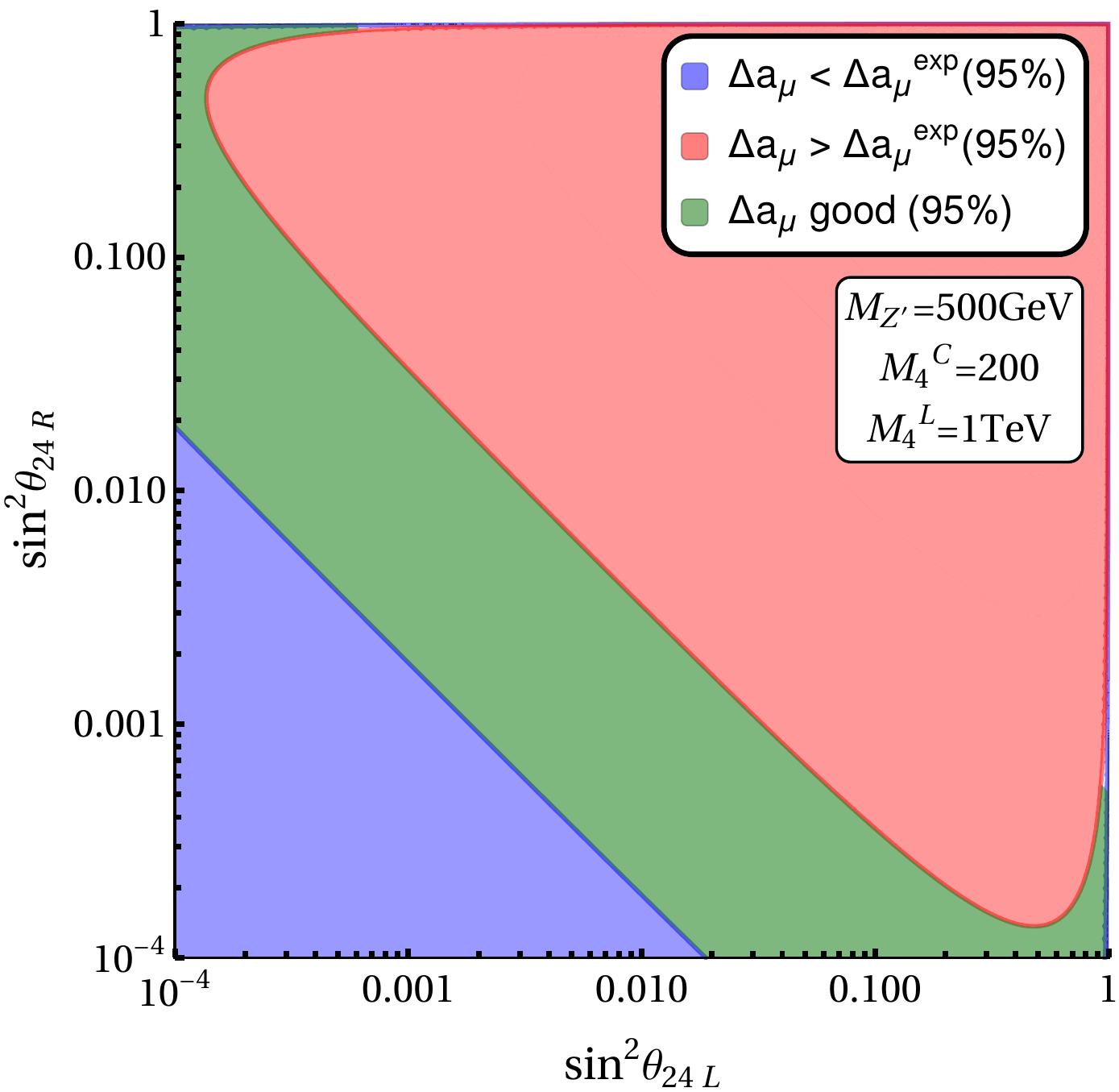}
		\captionsetup{width=0.8\linewidth}
		\caption{$\Delta a_\mu$ in angular parameter space with fixed $Z'$ mass}
		\label{fig:muong-2_and_trident2}
	\end{subfigure}
	\captionsetup{width=0.8\linewidth}
	\caption{Constraints in the $M_{Z'}$, $\sin^2\theta_{24L}$ and $\sin^2\theta_{24R}$ parameter space, mixing between the electron and vector-like lepton switched off.}
	\label{fig:muong-2_and_trident}
\end{figure}

Initially, we focus on the longest-standing anomaly, that of $(g-2)_\mu$. We first utilise a simple parameter space, as we require only mixing between the muon and vector-like lepton fields. To keep the analysis in a region potentially testable by upcoming future experiments, we take a vector-like fourth family lepton mass of $M_4^L=1\text{TeV}$ and a chirality-flipping fourth family mass of $M_4^C=200\text{GeV}$
(as discussed earlier we make a distinction between these two sources of mass). The smaller value of $M_4^C$
is well motivated by the need for perturbativity in Yukawa couplings, as the SM Higgs vev is 176GeV, since $M_4^C$ is proportional to the Higgs vev. For this investigation, the parameter space under test is detailed in Table  \ref{tab:parameters_for_muong-2_only_test}.
\begin{table}[H]
	\centering
	\renewcommand{\arraystretch}{1.2}
	\begin{tabular}{l|c|}
		\textbf{Parameter} & \textbf{Value/Scanned Region}\\
		\hline
		$M_{Z'}$ & $50\rightarrow1000$ GeV \\
		$M_4^C$ & $200$ GeV \\
		$M_4^L$ & $1000$ GeV \\
		$\sin^2\theta_{12L,R}$ & $0.0$ \\
		$\sin^2\theta_{14L}$ & $0.0$ \\
		$\sin^2\theta_{14R}$ & $0.0$ \\
		$\sin^2\theta_{24L,R}$ & $0.0\rightarrow 1.0$ \\
	\end{tabular}
	\caption{Explored parameter space for muon $g-2$ test.}
	\label{tab:parameters_for_muong-2_only_test}
\end{table}
Within the stated parameter space, expressions for the observables under test are simplified considerably, and with fixed $M_4^C$ and $M_4^L$ we constrain the space in terms of the three variables $\sin^2\theta_{24L}$, $\sin^2\theta_{24R}$ and $M_{Z'}$, as shown in Figure \ref{fig:muong-2_and_trident}. Note that, as $\theta_{12L,R}$ and $\theta_{14L,R}$ are set vanishing, contributions to $(g-2)_e$ and $\operatorname{BR}(\mu\rightarrow e\gamma)$ are necessarily vanishing, as can be readily seen from Equations \eqref{eqn:electron_g-2_contributions} and \eqref{eqn:mu_e_gamma}. The dominant contribution to $(g-2)_\mu$ under these assumptions is shown in the final Feynman diagram in Figure \ref{fig:muong2_feynman_diagrams}, that with the enhancement factor of $M_4^C/m_\mu$.
The legend in Figure \ref{fig:muong-2_and_trident} shows the constraint from neutrino trident production as `$\nu\text{Tri}$' for brevity. Using only mixing between the muon and the vector-like lepton, it is not possible to predict a value for the electron $g-2$ consistent with the observed value as the electron-$Z'$ coupling does not exist. In order to recover this, we must consider mixing of the vector-like lepton with the electron, detailed in the following subsection.

\subsection{Anomalous electron magnetic moment}
\begin{figure}[H]
	\centering
	\includegraphics[width=0.5\textwidth]{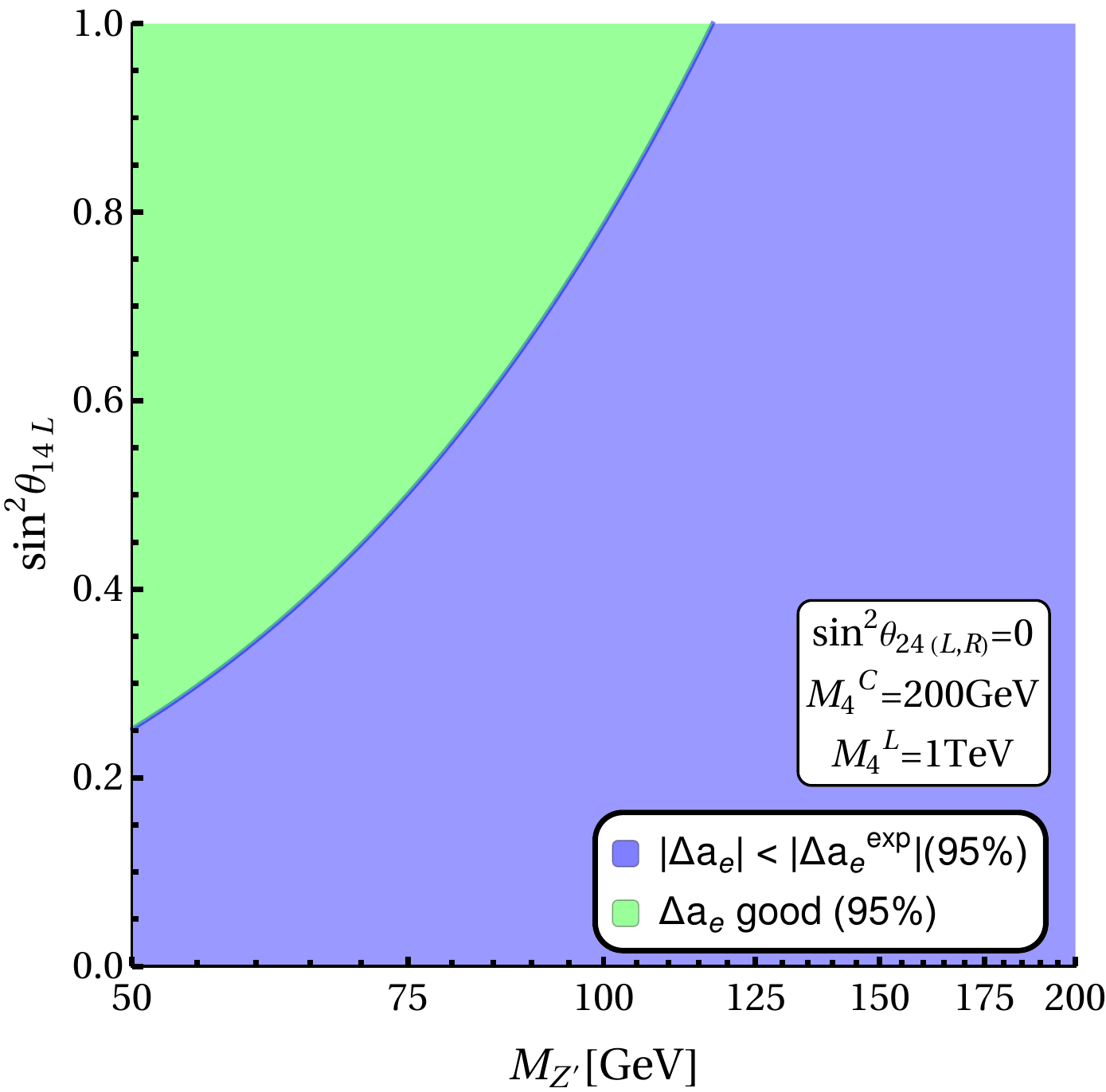}
	\captionsetup{width=0.8\linewidth}
	\caption{$\Delta a_e$ impact on $\sin^2\theta_{14L}$, $M_{Z'}$ parameter space, mixing between the muon and vector-like lepton switched off.}
	\label{fig:electron_g-2_only}
\end{figure}

Here we concentrate on the $(g-2)_e$. In order to test this observable alone, we investigate only mixing between the electron and vector-like lepton, and ignore any muon contributions. The region of parameter space under test is given in Table \ref{tab:parameters_for_electrong-2_only_test}, note also that mixing with the right-handed electron field is not required to obtain a good prediction.
\begin{table}[H]
	\centering
	\renewcommand{\arraystretch}{1.2}
	\begin{tabular}{l|c|}
		\textbf{Parameter} & \textbf{Value/Scanned Region}\\
		\hline
		$M_{Z'}$ & $50\rightarrow1000$ GeV \\
		$M_4^C$ & $200$ GeV \\
		$M_4^L$ & $1000$ GeV \\
		$\sin^2\theta_{12L,R}$ & $0.0$ \\
		$\sin^2\theta_{14L}$ & $0.0\rightarrow1.0$ \\
		$\sin^2\theta_{14R}$ & $0.0$ \\
		$\sin^2\theta_{24L,R}$ & $0.0$ \\
	\end{tabular}
	\caption{Explored parameter space for electron $g-2$ test.}
	\label{tab:parameters_for_electrong-2_only_test}
\end{table}
In Figure \ref{fig:electron_g-2_only}, we colour the electron $g-2$ being greater than the observed value (i.e. `less negative' than the experimental data) as the blue region, as such values are more SM-like. Blue regions therefore ameliorate the SM's tension with the experimental data but do not fully resolve it.
Similarly to the preceeding section, because there are no couplings between the electron and the muon (even at the loop level), there are no contributions to the CLFV decay $\mu\rightarrow e\gamma$. Similarly, there are no amendments to the SM expressions for the muon $g-2$ or neutrino trident decay. From this analysis one can conclude that only through using mixing between both muons and electrons with the vector-like leptons is it possible to simultaneously predict observed values of both the anomalous magnetic moments.

\subsection{Attempt to explain both anomalous moments}
In an attempt satisfy all constraints simultaneously, we set specific values for $M_{Z'}$, $M^C_4$ and $\sin^2\theta_{14}^L$ that inhabit allowed regions of parameter space in Figures \ref{fig:muong-2_and_trident1}, \ref{fig:muong-2_and_trident2} and \ref{fig:electron_g-2_only}, then scan through angular mixing parameters as before. The investigated region is summarised in Table \ref{tab:parameters_for_first_attempt_scan}. The choice of $Z'$ mass here is motivated by studying the regions of Figures \ref{fig:muong-2_and_trident} and \ref{fig:electron_g-2_only} that admit muon and electron $(g-2)$s respectively.
\begin{table}[H]
	\centering
	\renewcommand{\arraystretch}{1.2}
	\begin{tabular}{l|c|}
		\textbf{Parameter/Observable} & \textbf{Value/Scanned Region}\\
		\hline
		$M_{Z'}$ & $75$ GeV \\
		$M_4^C$ & $200$ GeV \\
		$M_4^L$ & $1000$ GeV \\
		$\sin^2\theta_{12L,R}$ & $0.0$ \\
		$\sin^2\theta_{14L}$ & $0.75$ \\
		$\sin^2\theta_{14R}$ & $0.0$ \\
		$\sin^2\theta_{24L,R}$ & $10^{-7}\rightarrow 1.0$ \\
		\hline
		$\operatorname{BR}(\mu\rightarrow e\gamma)$ & $10^{-3}\rightarrow 1.0$\\
	\end{tabular}
	\captionsetup{width=0.8\linewidth}
	\caption{Parameter space and $\operatorname{BR}(\mu\rightarrow e\gamma)$ in a parameter space where the electron and muon both mix with the vector-like lepton. Initial attempt to satisfy both anomalous moments.}
	\label{tab:parameters_for_first_attempt_scan}
\end{table}
This story concludes quite quickly with all points being excluded. The enchancement factor of $M^C_4/m_\mu$ in Equation \eqref{eqn:muon_g-2_contributions} is largely responsible for $(g-2)_\mu$ in this scenario, however such a term also gives an unacceptably large contribution to $\operatorname{BR}(\mu\rightarrow e\gamma)$ as per Equation \eqref{eqn:mu_e_gamma}, resulting in a branching fraction far above the experimental limit; the minimum $\operatorname{BR}(\mu\rightarrow e\gamma)$ for any parameter points in this scenario is around $10^{-3}$, as shown in Table \ref{tab:parameters_for_first_attempt_scan}. Such a situation persists even if $\sin^2\theta_{14}^L$ is scanned through it's entire range, and furthermore is unchanged by the choice of $M_4^L$, and is insensitive to the $Z'$ mass in the case of large $M_{4}^{C}$. We conclude therefore, that with a large chirality-flipping mass circa 200 GeV, it is not possible to simultaneously satisfy constraints and make predictions consistent with current data. This conclusion is consistent with the analytic arguments of the previous section, where the large contributions coming from large chirality flipping fourth family masses $M_4^C$ were assumed to dominate. We now go beyond this approximation, considering henceforth very small $M_4^C$.

If one sets $M_4^C$ vanishing, terms proportional to the aforementioned enhancement factor also vanish, eliminating the largest contribution to $\mu\rightarrow e\gamma$, as follows from Equation \eqref{eqn:mu_e_gamma}. Motivated by this reduction in the most restrictive decay the above analysis is repeated, but with the chirality-flipping mass removed.

\subsubsection{Vanishing $M^C_4$}
\begin{figure}[H]
	\centering
	\begin{subfigure}{0.48\textwidth}
		\includegraphics[width=1.0\textwidth]{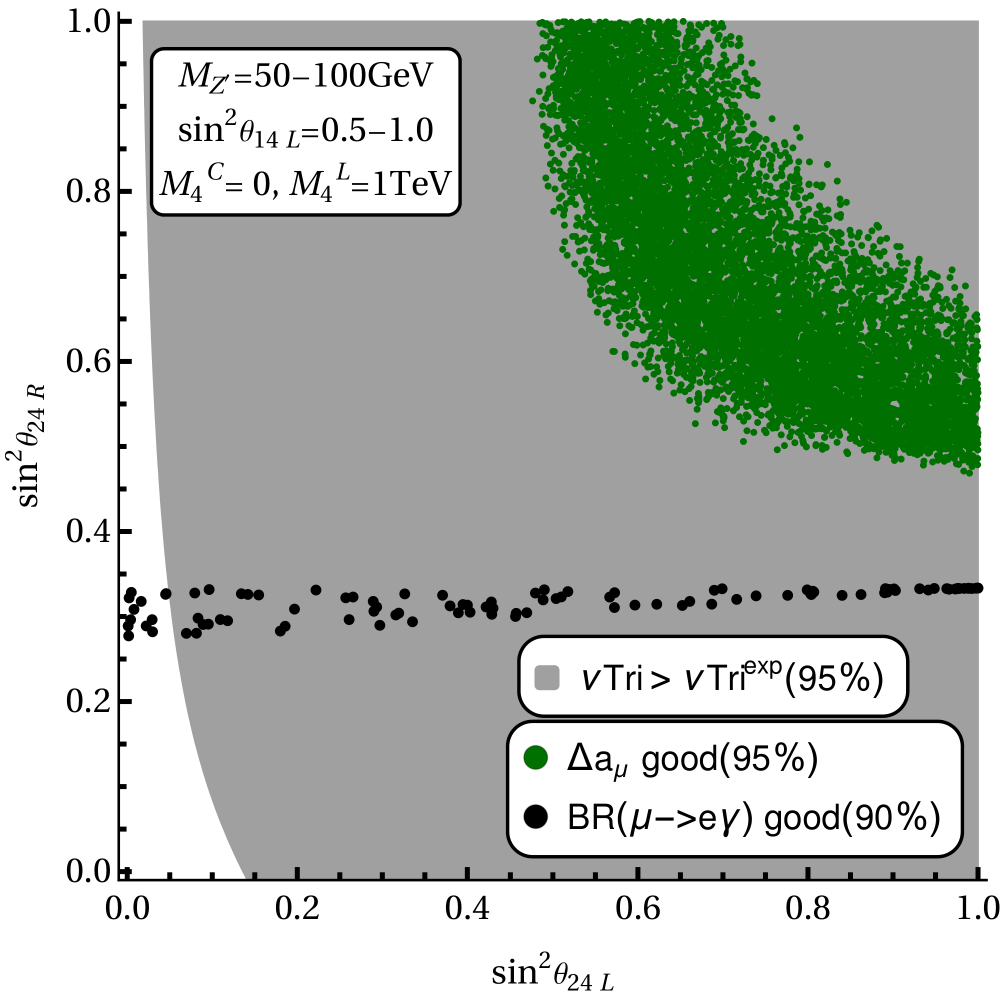}
		\captionsetup{width=0.8\textwidth}
		\caption{Parameter points that resolve $\Delta a_\mu$ and separately, points allowed under the $\mu\rightarrow e\gamma$ constraint. Fixed parameters given in legend. Chirality-flipping mass is set vanishing. All good $\Delta a_\mu$ points are excluded
		by trident and $\mu\rightarrow e\gamma$.}
		\label{fig:scan_without_M4C_g-2mu_and_muegamma}
	\end{subfigure}
	\begin{subfigure}{0.48\textwidth}
		\includegraphics[width=1.0\textwidth]{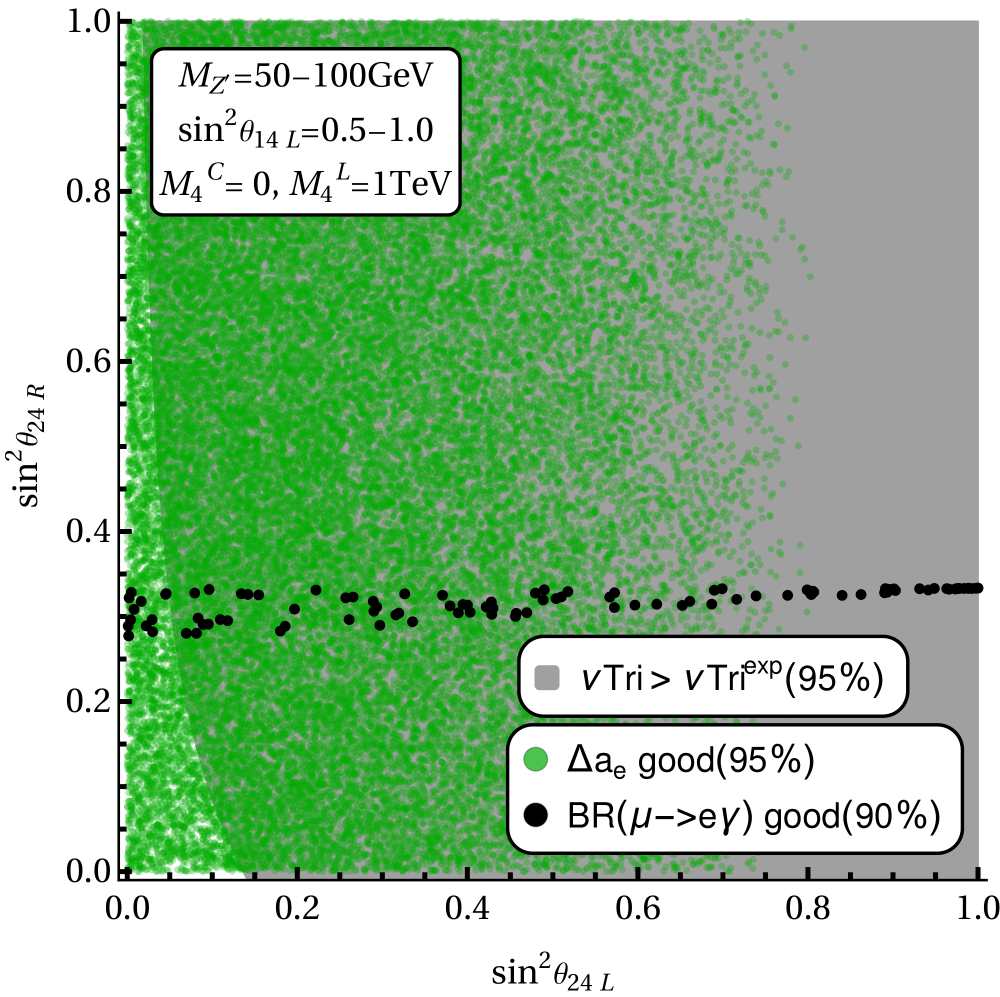}
		\captionsetup{width=0.8\textwidth}
		\caption{Parameter points that resolve $\Delta a_e$ and separately, points allowed under $\mu\rightarrow e\gamma$. Fixed parameters given in legend. Chirality-flipping mass is set vanishing.
		Some good $\Delta a_e$ points are allowed 
		by trident and $\mu\rightarrow e\gamma$.}
		\label{fig:scan_without_M4C_g-2e_and_muegamma}
	\end{subfigure}
	\captionsetup{width=0.8\textwidth}
	\caption{Parameter scan results for $M_4^C=0$.}
	\label{fig:scan_without_M4C_results}
\end{figure}

If we choose to turn off the chirality-flipping mass of the vector-like leptons, their mass becomes composed entirely of $M_4^L$. Terms proportional to the enhancement factor $M_4^C/m_\mu$ in Equation \eqref{eqn:muon_g-2_contributions} are sacrificed, which makes achieving a muon $g-2$ that is consistent with the experimental result more challenging. Larger mixing between the muon and vector-like leptons is required, but more freedom exists with respect to $BR(\mu\rightarrow e\gamma)$. We investigated a region of parameter space defined as per Table \ref{tab:parameters_for_large_scan_no_M4C}, to test its viability.
\begin{table}[H]
	\centering
	\renewcommand{\arraystretch}{1.2}
	\begin{tabular}{l|c|}
		\textbf{Parameter} & \textbf{Value/Scanned Region}\\
		\hline
		$M_{Z'}$ & $50\rightarrow 100$ GeV \\
		$M_4^C$ & $0$ GeV \\
		$M_4^L$ & $1000$ GeV \\
		$\sin^2\theta_{12L,R}$ & $0.0$ \\
		$\sin^2\theta_{14L}$ & $0.5\rightarrow 1.0$ \\
		$\sin^2\theta_{14R}$ & $0.0$ \\
		$\sin^2\theta_{24L,R}$ & $0.0\rightarrow 1.0$ \\
	\end{tabular}
	\caption{Parameters for scan without chirality-flipping mass.}
	\label{tab:parameters_for_large_scan_no_M4C}
\end{table}
For the results of this scan we consider the impact of each constraint separately, then check for overlap of allowed regions. Note that in Figure \ref{fig:scan_without_M4C_results}, angular parameters \emph{and} the heavy vector $Z^\prime$ mass are varied simultaneously, hence here we randomly select points and evaluate relevant phenomena, rather than excluding regions in the space. This also explains the spread of parameter points as compared to the previous exclusions.  Note that the range of $\sin^2\theta_{14}^L$ has been restricted in Tables \ref{tab:parameters_for_large_scan_no_M4C} and \ref{tab:parameters_for_large_scan_small_M4C} due to the fact that no points that satisfy $\operatorname{BR}(\mu\rightarrow e\gamma)$ could be found with $\sin^2\theta_{14}^L<0.5$, omitting this region increases the efficiency of our parameter scan. We also limit the ranges of $M_{Z'}$ in Tables \ref{tab:parameters_for_large_scan_no_M4C} and \ref{tab:parameters_for_large_scan_small_M4C} as $Z'$ masses much higher than this were found to be incompatible with $(g-2)_\mu$, and masses much below saturated the bound from $\mu\rightarrow e\gamma$.

In Figure \ref{fig:scan_without_M4C_g-2mu_and_muegamma}, one can see that, as suspected, larger $\sin^2\theta_{24L,R}$ mixings are required to obtain a muon $(g-2)_\mu$ consistent with current data. 
However, there is no overlapped region in Figure \ref{fig:scan_without_M4C_g-2mu_and_muegamma}, and $(g-2)_\mu$ cannot be solved without violating the muon decay constraint for a vanishing chirality-flipping mass, or the shown exclusion for neutrino trident production. On the other hand, Figure~\ref{fig:scan_without_M4C_g-2e_and_muegamma} shows that there are points that resolve the SM's tension with $(g-2)_e$, and are allowed by the strict $\operatorname{BR}(\mu\rightarrow e\gamma)$ limit and neutrino trident production. The lack of terms with the enhancement factor of $M^C_4/m_\mu$ in Equation \eqref{eqn:mu_e_gamma} means that points have been found with an acceptable branching fraction of $\mu\rightarrow e\gamma$ that was not possible with a large $M_4^C$.
 
Note that in both panels of Figure \ref{fig:scan_without_M4C_results} the most conservative neutrino trident limit is shown, where we assume that $M_{Z'}$ is fixed at 50GeV. We have also found that there is also no obvious correlation between $M_{Z'}$ and $\sin^2\theta_{14L}$ for $\mu\rightarrow e\gamma$, and points appear to be randomly distributed in this space. Since we have seen that neither large nor vanishing $M_4^C$ are viable, in the next subsection we switch on a small but non-zero $M_4^C$, to investigate if it may be possible to increase $(g-2)_\mu$ to an acceptable level, without giving an overlarge contribution to the CLFV muon decay.

\subsubsection{Small $M^C_4$ $\mathcal{O}(m_\mu)$}
\begin{figure}[H]
	\centering
	\begin{subfigure}{0.48\textwidth}
		\includegraphics[width=1.0\textwidth]{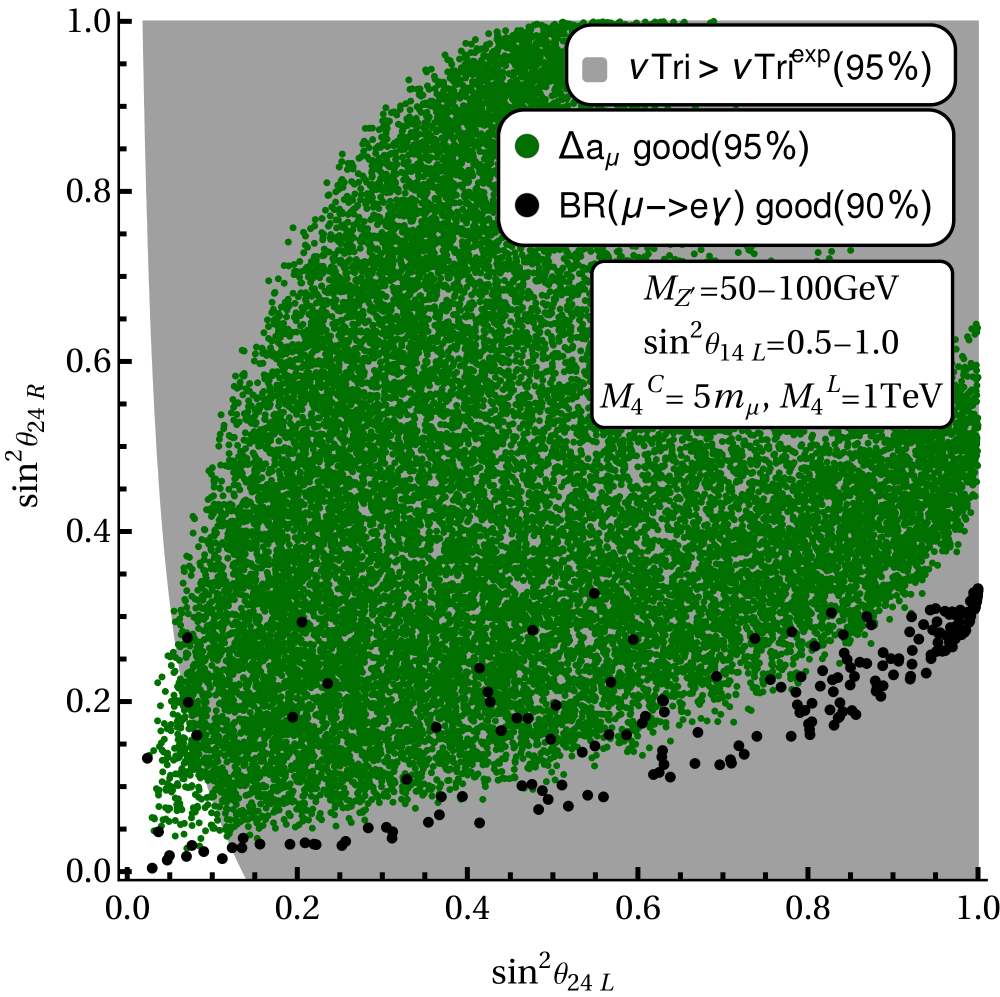}
		\caption{Parameter points that resolve $\Delta a_\mu$ and separately, points allowed under the $\mu\rightarrow e\gamma$ constraint. Fixed parameters given in legend, small chirality flipping mass. Unfortunately none of the points shown which have viable
		$\mu\rightarrow e\gamma$ and satisfy trident also have good $\Delta a_\mu$ (see text).
		}
		\label{fig:small_M4C_g-2mu_and_muegamma}
	\end{subfigure}
	\begin{subfigure}{0.48\textwidth}
		\includegraphics[width=1.0\textwidth]{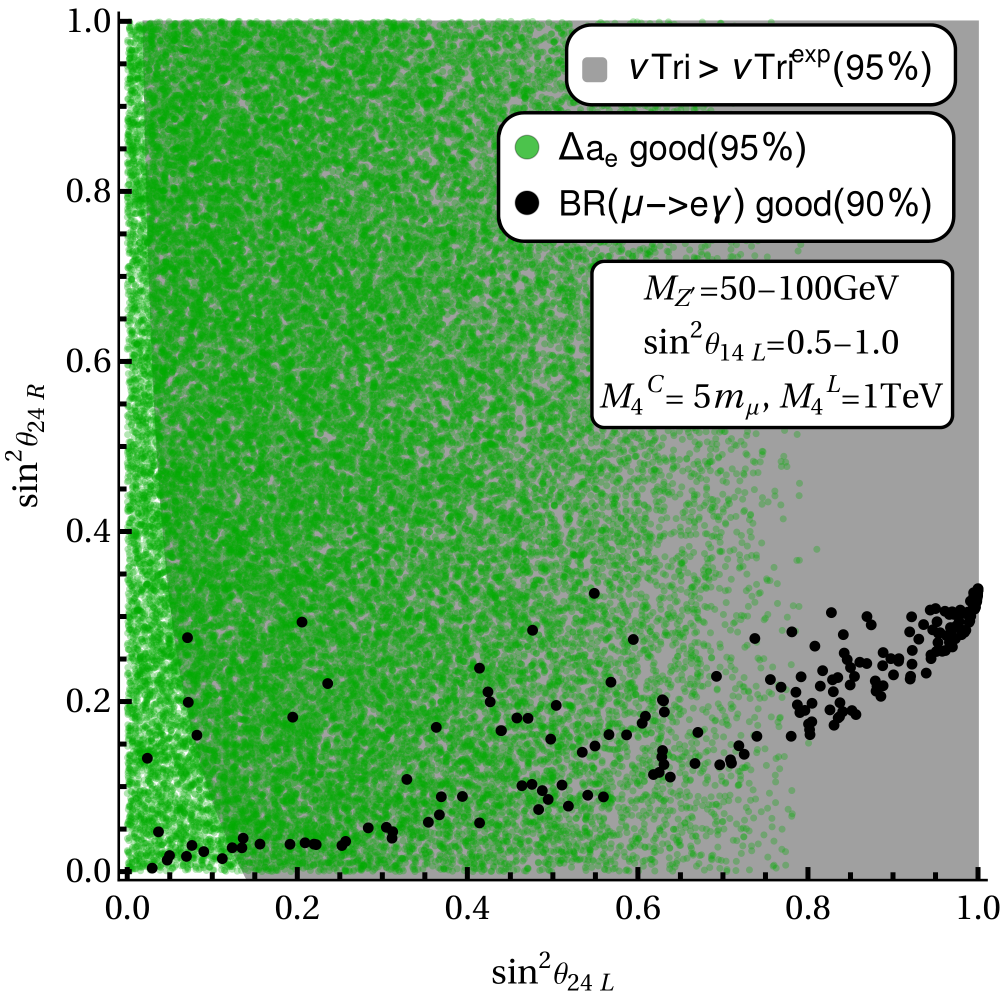}
		\captionsetup{width=0.8\textwidth}
		\caption{Parameter points that resolve $\Delta a_e$ and separately, points allowed under the $\mu\rightarrow e\gamma$ constraint. Fixed parameters given in legend, small chirality flipping mass.}
		\label{fig:small_M4C_g-2e_and_muegamma}
	\end{subfigure}
	\captionsetup{width=0.8\textwidth}
	\caption{Parameter scan results for small $M_4^C=5m_\mu$.}
	\label{fig:small_M4C_results}
\end{figure}

Here we perform analogous tests to those above but with a small chirality flipping mass, motivated by $(g-2)_\mu$ with the requirement that $\operatorname{BR}(\mu\rightarrow e\gamma)$ remains below the experimental limit. Ranges of parameters scanned in this investiagtion are given in Table \ref{tab:parameters_for_large_scan_small_M4C}.
\begin{table}[H]
	\centering
	\renewcommand{\arraystretch}{1.2}
	\begin{tabular}{l|c|}
		\textbf{Parameter} & \textbf{Value/Scanned Region}\\
		\hline
		$M_{Z'}$ & $50\rightarrow 100$ GeV \\
		$M_4^C$ & $5m_\mu$ \\
		$\sin^2\theta_{14L}$ & $0.5\rightarrow 1.0$ \\
		$\sin^2\theta_{14R}$ & $0.0$ \\
		$\sin^2\theta_{24L,R}$ & $0.0\rightarrow 1.0$ \\
		$\sin^2\theta_{12L,R}$ & $0.0$ \\
	\end{tabular}
	\caption{Parameters for larger scan with a small chirality-flipping mass.}
	\label{tab:parameters_for_large_scan_small_M4C}
\end{table}
Figure \ref{fig:small_M4C_results} shows points allowed under each separate observable in an analogous parameter space to Figure \ref{fig:scan_without_M4C_results}, but with $M_4^C=5m_\mu$. Once more neutrino trident production excludes a large region of the parameter space in this scenario. From initial study of the parameter space it seems that there is overlap between the allowed regions of $(g-2)_\mu$, $(g-2)_e$ and $\operatorname{BR}(\mu\rightarrow e\gamma)$, however, upon closer inspection of the parameter points allowed by $\mu\rightarrow e\gamma$, those points always yield negative (wrong sign) 
$(g-2)_\mu$ that is far away from the experimental value, and hence all points are excluded. 
In Table \ref{tab:view_allowed_points}, we examine more closely the points that are allowed under the most stringent constraint of  
$\mu\rightarrow e\gamma$. As 4th family mixing with the muons exists in this space, neutrino trident production is also a consideration, and the constraint of this observable in our space is given in Figure \ref{fig:small_M4C_results}. All points valid when considering $\operatorname{BR}(\mu\rightarrow e\gamma)$ exist with a small $\sin^2\theta_{24R}$ mixing angle, but can have a wide range of $Z'$ masses and $\sin^2\theta_{14L}$.
\begin{table}[H]
	\centering
	\renewcommand{\arraystretch}{1.3}
	\begin{tabular}{|l|l|l|l||l|l|l|}
		\hline
		\multicolumn{4}{|c||}{\textbf{Parameter}} & \multicolumn{3}{c|}{\textbf{Observable}} \\
		\hline
		$M_{Z'}/\text{GeV}$ & $\sin^2\theta_{14L}$ & $\sin^2\theta_{24L}$ & $\sin^2\theta_{24R}$ & $\operatorname{BR}(\mu\rightarrow e\gamma)$ & $\Delta a_e$ & $\Delta a_\mu$ \\
		\hline
		$69.5$ & $0.61$ & $0.11$ & $0.02$ & $3.25\times10^{-13}$ & $-2.15\times10^{-13}$ &  $-1.80\times10^{-10}$ \\
		$68.5$ & $0.80$ & $0.05$ & $0.01$ & $1.69\times10^{-13}$ & $-3.32\times10^{-13}$ & $-1.63\times10^{-10}$ \\
		$91.0$ & $0.99$ & $0.08$ & $0.16$ & $3.34\times10^{-13}$ & $-2.41\times10^{-13}$ & $-1.19\times10^{-9}$ \\
		$63.0$ & $0.99$ & $0.02$ & $0.13$ & $1.38\times10^{-13}$ & $-5.390\times10^{-13}$ & $-2.03\times10^{-9}$ \\
		$65.5$ & $0.78$ & $0.07$ & $0.02$ & $4.94\times10^{-14}$ & $-3.43\times10^{-13}$ & $-2.36\times10^{-10}$ \\
		$64.8$ & $0.78$ & $0.09$ & $0.02$ & $3.61\times10^{-13}$ & $-3.46\times10^{-13}$ & $-3.19\times10^{-10}$ \\
		$77.9$ & $0.85$ & $0.005$ & $0.02$ & $6.13\times10^{-14}$ & $-2.77\times10^{-13}$ & $-1.77\times10^{-10}$ \\
		$91.4$ & $0.81$ & $0.14$ & $0.04$ & $5.80\times10^{-14}$ & $-1.73\times10^{-13}$ & $-2.71\times10^{-10}$ \\
		$97.2$ & $0.86$ & $0.08$ & $0.03$ & $1.07\times10^{-13}$ & $-1.73\times10^{-13}$ & $-2.71\times10^{-10}$ \\
		$76.0$ & $0.63$ & $0.03$ & $0.004$ & $1.72\times10^{-13}$ & $-2.01\times10^{-13}$ & $-3.97\times10^{-11}$ \\
		$56.8$ & $0.96$ & $0.04$ & $0.05$ & $3.77\times10^{-14}$ & $-6.22\times10^{-13}$ & $-8.36\times10^{-10}$ \\
		$78.1$ & $0.99$ & $0.07$ & $0.20$ & $1.84\times10^{-14}$ & $-3.32\times10^{-13}$ & $-2.04\times10^{-9}$ \\
		$89.4$ & $1.0$ & $0.07$ & $0.28$ & $2.95\times10^{-13}$ & $-2.56\times10^{-13}$ & $-2.25\times10^{-9}$ \\
		\hline
	\end{tabular}
	\captionsetup{width=0.8\textwidth}
	\caption{Parameter points that are below the upper bound on $\operatorname{BR}(\mu\rightarrow e\gamma)$ for $M_4^C=5m_\mu$. The points in this table correspond to the 13 black points in Figure \ref{fig:small_M4C_results} that are also below the grey neutrino trident exclusion. These points do not satisfy the experimental value of $(\Delta a_{\mu })_{\exp}=(26.1\pm 8)\times10^{-10}$.}
	\label{tab:view_allowed_points}
\end{table}
We see that for the points in Table \ref{tab:view_allowed_points}, electron $g-2$ prefers regions of the space with small $\sin^2\theta_{24L}$, similarly to the preferred points under the neutrino trident constraint, given in the same plot as an excluded region derived in the same way as previous results for $M_4^C=0$. Many of these points are simultaneously consistent with the $\mu\rightarrow e \gamma$ limit, and also provide a $(g-2)_e$ consistent with experimental data (denoted in green), whilst a subset of these points do not violate the neutrino trident production limit. From these results, we can conclude that the best points lie in the region of small $\sin^2\theta_{24L}$ and $\sin^2\theta_{24R}$, and that such points simultaneously comply with $\operatorname{BR}(\mu\rightarrow e\gamma)$,  $(g-2)_e$ and neutrino trident. Such candidate points however do not allow for resolution of $\Delta a_\mu$, as they all have negative values for $\Delta a_\mu$, as opposed to the experimental value which is positive. 

A number of other chirality flipping masses were examined in this work, in the region $5m_\mu < M_4^C < 200\text{GeV}$, including a parameter scan whereby $M_4^C$ was randomly selected between these limits, and these tests yielded similar results to those shown in the last three sections, whereby it was not possible to obtain predictions that were simultaneously consistent with $(g-2)_e$, $(g-2)_\mu$ and $\operatorname{BR}(\mu\rightarrow e\gamma)$.

\section{Concluding Remarks}
\label{sec:conclusion}

In this paper, we have addressed the question: is it possible to explain the anomalous muon and electron $g-2$ in a $Z^{\prime }$ model? Although it is difficult to answer this question in general, since there are many possible $Z'$ models, we have seen that it is possible to consider a simple renormalisable and gauge invariant model in which the $Z'$ only has couplings to the electron and muon and their associated neutrinos, arising from mixing with a vector-like fourth family of leptons. This is achieved by assuming that only the vector-like leptons have non vanishing $U(1)'$ charges and are assumed to only mix with the first and second family of SM charged leptons. In this scenario, the heavy $Z^{\prime }$ gauge boson couples with the first and second family of SM charged leptons only through mixing with the vector-like generation. 

A feature of our analysis is to distinguish the two sources of mass for the 4th, vector-like family: 
the chirality flipping fourth family mass terms $M_4^C$ arising from the Higgs Yukawa couplings and are proportional to the Higgs vev
and the vector-like masses $M_{4}^{L}$ which are not proportional to the Higgs vev.
For the purposes of clarity we have
treated $M_4^C$ and $M_{4}^{L}$ as independent mass terms
in the analysis of the physical quantities of interest, rather than constructing the full fourth family mass matrix and 
diagonalising it, since such quantities rely on a chirality flip and are sensitive to $M_4^C$ rather than $M_{4}^{L}$. 

We began by assuming large fourth family chirality flipping masses
$M_4^C\gg m_\mu$, and showed that the expressions for $(g-2)_\mu$, $(g-2)_e$ and $\operatorname{BR}(\mu\rightarrow e\gamma)$
reduced to a minimal number of terms, all proportional to $M_4^C$. We were then able to construct an analytic argument which shows that 
it is not possible to explain the anomalous muon and electron $g-2$ in the $Z^{\prime }$ model, while respecting the 
bound on $\operatorname{BR}(\mu\rightarrow e\gamma)$.

We then performed a detailed numerical analysis of the parameter space of the above model, beginning with large 
$M_4^C =200$ GeV, where we showed that it is possible to account for $(g-2)_\mu$ in a region of parameter space
where the electron couplings were zero. Similarly, for $M_4^C =200$ GeV, we showed that it is possible to account for $(g-2)_e$ in a region of parameter space where the muon couplings were zero. In both cases $\operatorname{BR}(\mu\rightarrow e\gamma)$
was identically zero. 

Keeping $M_4^C =200$ GeV, we then attempted to explain both anomalous magnetic moments by switching on 
the couplings to the electron and muon simultaneously, but saw that it was not possible to do this while satisfying
$\operatorname{BR}(\mu\rightarrow e\gamma)$, as expected from the analytic arguments.

We then went beyond the regime of the analytic arguments by considering very small values of $M_4^C$.
With $M_4^C=0$, we saw that it is not possible to account for $(g-2)_\mu$ without violating the bounds from 
$\operatorname{BR}(\mu\rightarrow e\gamma)$ and trident, however it is possible to account for $(g-2)_e$
while respecting all constraints. With small but non-zero $M_4^C$ we reached similar conclusions, although the analysis
was more complicated, and it was necessary to examine specific benchmark points to reach this conclusion.

We stress that the fermiophobic
$Z'$ model is a good candidate to explain either $(g-2)_\mu$ or $(g-2)_e$, consistently with 
$\operatorname{BR}(\mu\rightarrow e\gamma)$ and trident, with the choice determined by the specific mixing scenario. 
However to explain the $(g-2)_\mu$ always requires a significant non-vanishing 
chirality flipping mass involving the 4th vector-like family of leptons.

We would like to comment on the generality of our conclusion that, for the $Z^\prime$ framework considered in this paper, we cannot simultaneously explain the electron and muon g-2 results within the relevant parameter space of the model,
while satisfying the constraints of $BR(\mu\rightarrow e\gamma)$ and neutrino trident production.
Does this conclusion apply to all $Z^\prime$ models? While it is impossible to answer this question absolutely, 
there are reasons why our results here might be considered very general and indicative of a large class
of $Z^\prime$ models. The main reason for this is that, in the considered 
framework, the $Z^\prime$ is only allowed to couple to the electron and muon and their associated 
neutrinos, arising from mixing with a vector-like fourth family of leptons, thereby 
eliminating the quark couplings and allowing us to focus on 
the connection between CLUV, CLFV and the electron and muon $g-2$ anomalies only, independently of other constraints.
Moreover, the allowed $Z^\prime$ couplings are free parameters in our approach and so may represent the couplings in a large class
of $Z^\prime$ models. Furthermore, we have presented a general analytic argument that provides some insight into our numerical
results. For example, we do not require the $Z^\prime$ to couple identically to left- and right-handed leptons, and the masses for intermediate particles in the one-loop diagrams cancel in the final expression for $\operatorname{BR}(\mu\rightarrow e\gamma)$ in Equation \ref{eqn:really_simplified_mu_e_gamma_expression}, which lends this result some generality. 
We also note that this paper represents the first paper to attempt to explain both electron and muon $g-2$ anomalies simultaneously
within a $Z'$ model. Thus, although the problem of the CLFV constraint in preventing an explanation of electron and muon $g-2$ anomalies is well known in general, it had not been studied within the framework of $Z^\prime$ models before the present paper. Indeed this is the first work we know of that attempts to explain the muon and electron anomalous magnetic moments simultaneously using a simple $Z^\prime$ model.

Finally we comment that since 
there are models in the literature which account for all these observables based on having scalars, it might be interesting to extend
the scalar sector of a $Z'$ model. 
The lepton flavour violating processes could then be used to set constraints on the masses for the CP even and CP odd heavy neutral scalars, as in \cite{CarcamoHernandez:2019xkb}. 
However, such a study 
is beyond the scope of the present paper. 

In conclusion, within a model where the $Z'$ only has tunable couplings to the electron and muon and their associated 
neutrinos, arising from mixing with a vector-like fourth family of leptons,
it is not possible to simultaneously satisfy the experimentally observed values of 
$(g-2)_\mu$ and $(g-2)_e$, while respecting the $\operatorname{BR}(\mu\rightarrow e\gamma)$
and trident constraints, within any of the exhaustively explored parameter space
(only one or other of $(g-2)_\mu$ or $(g-2)_e$ can be explained). 
Since the model allows complete freedom in the choice of couplings, and the diagrams involving 
fourth family lepton exchange can be chosen to contribute or not,
this model may be regarded as indicative of any $Z'$ model with gauge coupling and charges of order one.

\section*{Acknowledgments}
This research has received funding from Fondecyt (Chile), Grants
No.~1170803, CONICYT PIA/Basal FB0821. SFK acknowledges the STFC
Consolidated Grant ST/L000296/1 and the European Union's Horizon 2020
Research and Innovation programme under Marie Sk\l {}odowska-Curie grant
agreements Elusives ITN No.\ 674896 and InvisiblesPlus RISE No.\ 690575.
AECH thanks the University of Southampton, where this work was
started, for its hospitality. SJR is supported by a Mayflower studentship at the University of
Southampton.


\appendix
\section{Further Analytics for Observables}
It is important to understand how the observables $\operatorname{BR} ( \mu \rightarrow e \gamma)$, muon $g-2$, electron $g-2$ and neutrino trident can be written in terms of the mixing angles. The coupling constants appearing in each observable consist of the mixing angles. The coupling constants are defined from Equation \eqref{eqn:mu_mu_zprime_coupling} to \eqref{eqn:mu_e_zprime_coupling} in Section \ref{sec:LFV_in_this_model}. 

\subsection{The branching ratio of $\mu \rightarrow e \gamma$}

The branching ratio of $\mu \rightarrow e \gamma$ is the following:
\begin{equation}
\operatorname{BR}(\mu\rightarrow e\gamma) = \frac{\alpha}{1024 \pi^4} \frac{m_{\mu}^5}{M_{Z^{\prime}}^4 \Gamma_\mu} (\left\vert \widetilde{\sigma }_{L}\right\vert ^{2}+\left\vert \widetilde{\sigma }_{R}\right\vert^{2})
\label{eqn:analytic_expression_for_muegamma_1}
\end{equation}
The $\widetilde{\sigma}_{L,R}$ are given by:
\begin{align}
\begin{split}
\widetilde{\sigma }_{L}& =\sum_{a=e,\mu,E}\left[(g_{L})_{ea}(g_{L})_{a\mu}F(x_{a})+\frac{m_{a}}{m_{\mu }}(g_{L})_{ea}(g_{R})_{a\mu}G(x_{a})\right] , \\
\widetilde{\sigma }_{R}& =\sum_{a=e,\mu,E}\left[ (g_{R})_{ea}(g_{R})_{a\mu}F(x_{a})+\frac{m_{a}}{m_{\mu }}(g_{R})_{ea}(g_{L})_{a\mu}G(x_{a})\right],\hspace{1.5cm}x_{a}=\frac{m_{a}^{2}}{M_{Z^{\prime }}^{2}}
\label{eqn:contributions_to_muegamma_(sigmas)_1}
\end{split}
\end{align}
Expanding the above $\widetilde{\sigma}_{L,R}$ in terms of electron, muon and fourth family:
\begin{align}
\begin{split}
\widetilde{\sigma}_L = \Big[ &\left( g_L \right)_{ee} \left( g_L \right)_{e\mu} F\left( x_e \right) + \frac{m_e}{m_\mu}\left( g_L \right)_{ee} \left( g_R \right)_{e\mu} G\left( x_e \right) \\
&\left( g_L \right)_{e\mu} \left( g_L \right)_{\mu\mu} F\left( x_\mu \right) + \frac{m_\mu}{m_\mu}\left( g_L \right)_{e\mu} \left( g_R \right)_{\mu\mu} G\left( x_\mu \right) \\
&\left( g_L \right)_{eE} \left( g_L \right)_{E\mu} F\left( x_{E} \right) + \frac{M_4^C}{m_\mu}\left( g_L \right)_{eE} \left( g_R \right)_{E\mu} G\left( x_{E} \right) \Big] \\
\widetilde{\sigma}_R = \Big[ &\left( g_R \right)_{ee} \left( g_R \right)_{e\mu} F\left( x_e \right) + \frac{m_e}{m_\mu}\left( g_R \right)_{ee} \left( g_L \right)_{e\mu} G\left( x_e \right) \\
&\left( g_R \right)_{e\mu} \left( g_R \right)_{\mu\mu} F\left( x_\mu \right) + \frac{m_\mu}{m_\mu}\left( g_R \right)_{e\mu} \left( g_L \right)_{\mu\mu} G\left( x_\mu \right) \\
&\left( g_R \right)_{eE} \left( g_R \right)_{E\mu} F\left( x_{E} \right) + \frac{M_4^C}{m_\mu}\left( g_R \right)_{eE} \left( g_L \right)_{E\mu} G\left( x_{E} \right) \Big]
\label{eqn:expansion_of_sigmas_with_each_family}
\end{split}
\end{align}
One important feature in Equation \eqref{eqn:expansion_of_sigmas_with_each_family} is the chirality-flipping mass was used instead of vector-like mass in the last line of Equation \eqref{eqn:expansion_of_sigmas_with_each_family}. It then is possible to turn the coupling constants in each $\widetilde{\sigma}$ into the mixing angles by using the Equations \eqref{eqn:mu_mu_zprime_coupling}-\eqref{eqn:mu_e_zprime_coupling}. It was assumed that $g^{\prime} q_{L4}$ in each coupling constant to be $1$.
\begin{align}
\begin{split}
\widetilde{\sigma}_L = \Big[ &\Big(\sin\theta_{12}^{L}\sin\theta_{24}^{L}+\cos\theta_{12}^{L}\cos\theta_{24}^{L}\sin\theta_{14}^{L}\Big)^2 \times \\
&\Big( \sin\theta_{12}^{L}\sin\theta_{24}^{L}+\cos\theta_{12}^{L}\cos\theta_{24}^{L}\sin\theta_{14}^{L}\Big)\Big(\cos\theta_{12}^{L}\sin\theta_{24}^{L}-\cos\theta_{24}^{L}\sin\theta_{12}^{L}\sin\theta_{14}^{L} \Big) F\left( x_1 \right) \\
& + \frac{m_1}{m_2} \Big(\sin\theta_{12}^{L}\sin\theta_{24}^{L}+\cos\theta_{12}^{L}\cos\theta_{24}^{L}\sin\theta_{14}^{L}\Big)^2 \times \\
& \Big( \sin\theta_{12}^{R}\sin\theta_{24}^{R}+\cos\theta_{12}^{R}\cos\theta_{24}^{R}\sin\theta_{14}^{R}\Big)\Big(\cos\theta_{12}^{R}\sin\theta_{24}^{R}-\cos\theta_{24}^{R}\sin\theta_{12}^{R}\sin\theta_{14}^{R} \Big) G\left( x_1 \right) \\
& + \Big( \sin\theta_{12}^{L}\sin\theta_{24}^{L}+\cos\theta_{12}^{L}\cos\theta_{24}^{L}\sin\theta_{14}^{L}\Big)\Big(\cos\theta_{12}^{L}\sin\theta_{24}^{L}-\cos\theta_{24}^{L}\sin\theta_{12}^{L}\sin\theta_{14}^{L} \Big) \times \\
& \Big(\cos\theta_{12}^{L}\sin\theta_{24}^{L}-\cos\theta_{24}^{L}\sin\theta_{12}^{L}\sin\theta_{14}^{L}\Big)^2 F\left( x_2 \right) \\
& + \frac{m_2}{m_2} \Big( \sin\theta_{12}^{L}\sin\theta_{24}^{L}+\cos\theta_{12}^{L}\cos\theta_{24}^{L}\sin\theta_{14}^{L}\Big)\Big(\cos\theta_{12}^{L}\sin\theta_{24}^{L}-\cos\theta_{24}^{L}\sin\theta_{12}^{L}\sin\theta_{14}^{L} \Big) \times \\
& \Big(\cos\theta_{12}^{R}\sin\theta_{24}^{R}-\cos\theta_{24}^{R}\sin\theta_{12}^{R}\sin\theta_{14}^{R}\Big)^2 G\left( x_2 \right) \\	
& + \cos\theta_{14}^{L}\cos\theta_{24}^{L} \Big(\sin\theta_{12}^{L}\sin\theta_{24}^{L}+\cos\theta_{12}^{L}\cos\theta_{24}^{L}\sin\theta_{14}^{L}\Big) \times \\
& \cos\theta_{14}^{L}\cos\theta_{24}^{L}\Big(\cos\theta_{12}^{L}\sin\theta_{24}^{L}-\cos\theta_{24}^{L}\sin\theta_{12}^{L}\sin\theta_{14}^{L}\Big) F\left( x_4 \right) \\
& + \frac{M_4^C}{m_2} \cos\theta_{14}^{L}\cos\theta_{24}^{L} \Big(\sin\theta_{12}^{L}\sin\theta_{24}^{L}+\cos\theta_{12}^{L}\cos\theta_{24}^{L}\sin\theta_{14}^{L}\Big) \times \\
& \cos\theta_{14}^{R}\cos\theta_{24}^{R}\Big(\cos\theta_{12}^{R}\sin\theta_{24}^{R}-\cos\theta_{24}^{R}\sin\theta_{12}^{R}\sin\theta_{14}^{R}\Big) G\left( x_4 \right) \Big] \\	
\widetilde{\sigma}_R = \Big[ &\Big(\sin\theta_{12}^{R}\sin\theta_{24}^{R}+\cos\theta_{12}^{R}\cos\theta_{24}^{R}\sin\theta_{14}^{R}\Big)^2 \times \\
&\Big( \sin\theta_{12}^{R}\sin\theta_{24}^{R}+\cos\theta_{12}^{R}\cos\theta_{24}^{R}\sin\theta_{14}^{R}\Big)\Big(\cos\theta_{12}^{R}\sin\theta_{24}^{R}-\cos\theta_{24}^{R}\sin\theta_{12}^{R}\sin\theta_{14}^{R} \Big) F\left( x_1 \right) \\
& + \frac{m_1}{m_2} \Big(\sin\theta_{12}^{R}\sin\theta_{24}^{R}+\cos\theta_{12}^{R}\cos\theta_{24}^{R}\sin\theta_{14}^{R}\Big)^2 \times \\
& \Big( \sin\theta_{12}^{L}\sin\theta_{24}^{L}+\cos\theta_{12}^{L}\cos\theta_{24}^{L}\sin\theta_{14}^{L}\Big)\Big(\cos\theta_{12}^{L}\sin\theta_{24}^{L}-\cos\theta_{24}^{L}\sin\theta_{12}^{L}\sin\theta_{14}^{L} \Big) G\left( x_1 \right) \\
& + \Big( \sin\theta_{12}^{R}\sin\theta_{24}^{R}+\cos\theta_{12}^{R}\cos\theta_{24}^{R}\sin\theta_{14}^{R}\Big)\Big(\cos\theta_{12}^{R}\sin\theta_{24}^{R}-\cos\theta_{24}^{R}\sin\theta_{12}^{R}\sin\theta_{14}^{R} \Big) \times \\
& \Big(\cos\theta_{12}^{R}\sin\theta_{24}^{R}-\cos\theta_{24}^{R}\sin\theta_{12}^{R}\sin\theta_{14}^{R}\Big)^2 F\left( x_2 \right) \\
& + \frac{m_2}{m_2} \Big( \sin\theta_{12}^{R}\sin\theta_{24}^{R}+\cos\theta_{12}^{R}\cos\theta_{24}^{R}\sin\theta_{14}^{R}\Big)\Big(\cos\theta_{12}^{R}\sin\theta_{24}^{R}-\cos\theta_{24}^{R}\sin\theta_{12}^{R}\sin\theta_{14}^{R} \Big) \times \\
& \Big(\cos\theta_{12}^{L}\sin\theta_{24}^{L}-\cos\theta_{24}^{L}\sin\theta_{12}^{L}\sin\theta_{14}^{L}\Big)^2 G\left( x_2 \right) \\	
& + \cos\theta_{14}^{R}\cos\theta_{24}^{R} \Big(\sin\theta_{12}^{R}\sin\theta_{24}^{R}+\cos\theta_{12}^{R}\cos\theta_{24}^{R}\sin\theta_{14}^{R}\Big) \times \\
& \cos\theta_{14}^{R}\cos\theta_{24}^{R}\Big(\cos\theta_{12}^{R}\sin\theta_{24}^{R}-\cos\theta_{24}^{R}\sin\theta_{12}^{R}\sin\theta_{14}^{R}\Big) F\left( x_4 \right) \\
& + \frac{M_4^C}{m_2} \cos\theta_{14}^{R}\cos\theta_{24}^{R} \Big(\sin\theta_{12}^{R}\sin\theta_{24}^{R}+\cos\theta_{12}^{R}\cos\theta_{24}^{R}\sin\theta_{14}^{R}\Big) \times \\
& \cos\theta_{14}^{L}\cos\theta_{24}^{L}\Big(\cos\theta_{12}^{L}\sin\theta_{24}^{L}-\cos\theta_{24}^{L}\sin\theta_{12}^{L}\sin\theta_{14}^{L}\Big) G\left( x_4 \right) \Big] \\	
\label{eqn:expansion_of_sigmas_with_mixing_angles}
\end{split}
\end{align}

\subsection{Anomalous muon $g-2$}

The anomalous muon $g-2$ is given by:
\begin{equation}
\Delta a_{\mu}^{Z^{\prime}} = -\frac{m_{\mu}^2}{8 \pi^2 M_{Z^{\prime}}^2} \sum_{a=e,\mu,E} \left[ \left( \lvert \left( g_L \right)_{\mu a} \rvert^2 + \lvert \left( g_R \right)_{\mu a} \rvert^2 \right) F(x_a) + \frac{m_a}{m_\mu} \operatorname{Re} \left[ \left( g_L \right)_{\mu a} \left( g_R^{*} \right)_{\mu a} \right] G(x_a) \right],\hspace{1cm}x_{a}=\frac{m_{a}^{2}}{M_{Z^{\prime }}^{2}}.
\label{eqn:analytic_expression_for_muong2}
\end{equation}
Expanding the above equation in terms of electron, muon and vector-like lepton couplings as per $\operatorname{BR}\left(\mu \rightarrow e \gamma \right)$:
\begin{align}
\begin{split}
\Delta a_{\mu}^{Z^{\prime}} = -\frac{m_{\mu}^2}{8 \pi^2 M_{Z^{\prime}}^2} \bigg[ &\left( \lvert \left( g_L \right)_{\mu e} \rvert^2 + \lvert \left( g_R \right)_{\mu e} \rvert^2 \right) F(x_e) + \frac{m_e}{m_\mu} \operatorname{Re} \left[ \left( g_L \right)_{\mu e} \left( g_R^{*} \right)_{\mu e} \right] G(x_e) \\
&+ \left( \lvert \left( g_L \right)_{\mu \mu} \rvert^2 + \lvert \left( g_R \right)_{\mu \mu} \rvert^2 \right) F(x_\mu) + \frac{m_\mu}{m_\mu} \operatorname{Re} \left[ \left( g_L \right)_{\mu \mu} \left( g_R^{*} \right)_{\mu \mu} \right] G(x_\mu) \\
&+ \left( \lvert \left( g_L \right)_{\mu E} \rvert^2 + \lvert \left( g_R \right)_{\mu E} \rvert^2 \right) F(x_{E}) + \frac{M_4^C}{m_\mu} \operatorname{Re} \left[ \left( g_L \right)_{\mu E} \left( g_R^{*} \right)_{\mu E} \right] G(x_{E}) \bigg]
\label{eqn:expanded_expression_for_muong2}
\end{split}
\end{align}
The chirality-flipping mass is used in the last line of equation \eqref{eqn:expanded_expression_for_muong2} similarly to Equation \eqref{eqn:expansion_of_sigmas_with_each_family}. It then is possible to represent $\Delta a_\mu$ in terms of mixing angles.
\begin{align}
\begin{split}
\Delta a_{\mu}^{Z^{\prime}} = -\frac{m_{\mu}^2}{8 \pi^2 M_{Z^{\prime}}^2} \bigg[ & \bigg( \lvert \Big(\sin\theta_{12}^{L}\sin\theta_{24}^{L}+\cos\theta_{12}^{L}\cos\theta_{24}^{L}\sin\theta_{14}^{L}\Big)\Big(\cos\theta_{12}^{L}\sin\theta_{24}^{L}-\cos\theta_{24}^{L}\sin\theta_{12}^{L}\sin\theta_{14}^{L}\Big) \rvert^2 \\
&+ \lvert \Big(\sin\theta_{12}^{R}\sin\theta_{24}^{R}+\cos\theta_{12}^{R}\cos\theta_{24}^{R}\sin\theta_{14}^{R}\Big)\Big(\cos\theta_{12}^{R}\sin\theta_{24}^{R}-\cos\theta_{24}^{R}\sin\theta_{12}^{R}\sin\theta_{14}^{R}\Big) \rvert^2 \bigg) F(x_1) \\
&+ \frac{m_1}{m_2} \Big(\sin\theta_{12}^{L}\sin\theta_{24}^{L}+\cos\theta_{12}^{L}\cos\theta_{24}^{L}\sin\theta_{14}^{L}\Big)\Big(\cos\theta_{12}^{L}\sin\theta_{24}^{L}-\cos\theta_{24}^{L}\sin\theta_{12}^{L}\sin\theta_{14}^{L}\Big) \\
&\times \Big(\sin\theta_{12}^{R}\sin\theta_{24}^{R}+\cos\theta_{12}^{R}\cos\theta_{24}^{R}\sin\theta_{14}^{R}\Big)\Big(\cos\theta_{12}^{R}\sin\theta_{24}^{R}-\cos\theta_{24}^{R}\sin\theta_{12}^{R}\sin\theta_{14}^{R}\Big) G(x_1) \\	
&+ \bigg( \lvert \Big(\cos\theta_{12}^{L}\sin\theta_{24}^{L}-\cos\theta_{24}^{L}\sin\theta_{12}^{L}\sin\theta_{14}^{L}\Big)^2 \rvert^2 + \lvert \Big(\cos\theta_{12}^{R}\sin\theta_{24}^{R}-\cos\theta_{24}^{R}\sin\theta_{12}^{R}\sin\theta_{14}^{R}\Big)^2 \rvert^2 \bigg) F(x_2) \\
&+ \frac{m_2}{m_2} \Big(\cos\theta_{12}^{L}\sin\theta_{24}^{L}-\cos\theta_{24}^{L}\sin\theta_{12}^{L}\sin\theta_{14}^{L}\Big)^2 \Big(\cos\theta_{12}^{R}\sin\theta_{24}^{R}-\cos\theta_{24}^{R}\sin\theta_{12}^{R}\sin\theta_{14}^{R}\Big)^2 G(x_2) \\
&+ \bigg( \lvert \cos\theta_{14}^{L}\cos\theta_{24}^{L}\Big(\cos\theta_{12}^{L}\sin\theta_{24}^{L}-\cos\theta_{24}^{L}\sin\theta_{12}^{L}\sin\theta_{14}^{L}\Big) \rvert^2 \\
&+ \lvert \cos\theta_{14}^{R}\cos\theta_{24}^{R}\Big(\cos\theta_{12}^{R}\sin\theta_{24}^{R}-\cos\theta_{24}^{R}\sin\theta_{12}^{R}\sin\theta_{14}^{R}\Big) \rvert^2 \bigg) F(x_4) \\
&+ \frac{M_4^C}{m_2} \cos\theta_{14}^{L}\cos\theta_{24}^{L}\Big(\cos\theta_{12}^{L}\sin\theta_{24}^{L}-\cos\theta_{24}^{L}\sin\theta_{12}^{L}\sin\theta_{14}^{L}\Big) \\
&\times \cos\theta_{14}^{R}\cos\theta_{24}^{R}\Big(\cos\theta_{12}^{R}\sin\theta_{24}^{R}-\cos\theta_{24}^{R}\sin\theta_{12}^{R}\sin\theta_{14}^{R}\Big) G(x_4)
\label{eqn:expanded_expression_for_muong2_mixing_angles}
\end{split}
\end{align}

\subsection{Anomalous electron $g-2$}

The anomalous electron $g-2$ is given by:
\begin{equation}
\Delta a_{e}^{Z^{\prime}} = -\frac{m_{e}^2}{8 \pi^2 M_{Z^{\prime}}^2} \sum_{a=e,\mu,E} \left[ \left( \lvert \left( g_L \right)_{ea} \rvert^2 + \lvert \left( g_R \right)_{ea} \rvert^2 \right) F(x_a) + \frac{m_a}{m_e} \operatorname{Re} \left[ \left( g_L \right)_{ea} \left( g_R^{*} \right)_{ea} \right] G(x_a) \right],\hspace{1cm}x_{a}=\frac{m_{a}^{2}}{M_{Z^{\prime }}^{2}}.
\label{eqn:analytic_expression_for_electrong2}
\end{equation}
Expanding the above equation in terms of electron, muon and vector-like lepton as previously, the form is
\begin{align}
\begin{split}
\Delta a_{e}^{Z^{\prime}} = -\frac{m_{e}^2}{8 \pi^2 M_{Z^{\prime}}^2} \bigg[ &\left( \lvert \left( g_L \right)_{ee} \rvert^2 + \lvert \left( g_R \right)_{ee} \rvert^2 \right) F(x_e) + \frac{m_e}{m_e} \operatorname{Re} \left[ \left( g_L \right)_{ee} \left( g_R^{*} \right)_{ee} \right] G(x_e) \\
&+ \left( \lvert \left( g_L \right)_{e\mu} \rvert^2 + \lvert \left( g_R \right)_{e\mu} \rvert^2 \right) F(x_\mu) + \frac{m_\mu}{m_e} \operatorname{Re} \left[ \left( g_L \right)_{e\mu} \left( g_R^{*} \right)_{e\mu} \right] G(x_\mu) \\
&+ \left( \lvert \left( g_L \right)_{eE} \rvert^2 + \lvert \left( g_R \right)_{eE} \rvert^2 \right) F(x_{E}) + \frac{M_4^C}{m_e} \operatorname{Re} \left[ \left( g_L \right)_{eE} \left( g_R^{*} \right)_{eE} \right] G(x_{E}) \bigg]
\label{eqn:expanded_expression_for_electrong2}
\end{split}
\end{align}
The chirality-flipping mass is used in the last line of Equation \eqref{eqn:expanded_expression_for_electrong2} similarly to the Equations \eqref{eqn:expansion_of_sigmas_with_each_family} or \eqref{eqn:expanded_expression_for_muong2}. It then is possible to represent anomalous electron $g-2$ in terms of mixing angles.
\begin{align}
\begin{split}
\Delta a_{e}^{Z^{\prime}} = -\frac{m_{e}^2}{8 \pi^2 M_{Z^{\prime}}^2} \bigg[ & \bigg( \lvert \Big(\sin\theta_{12}^{L}\sin\theta_{24}^{L}+\cos\theta_{12}^{L}\cos\theta_{24}^{L}\sin\theta_{14}^{L}\Big)^2 \rvert^2 + \lvert \Big(\sin\theta_{12}^{R}\sin\theta_{24}^{R}+\cos\theta_{12}^{R}\cos\theta_{24}^{R}\sin\theta_{14}^{R}\Big)^2 \rvert^2 \bigg) F(x_1) \\
&+ \frac{m_1}{m_1} \Big(\sin\theta_{12}^{L}\sin\theta_{24}^{L}+\cos\theta_{12}^{L}\cos\theta_{24}^{L}\sin\theta_{14}^{L}\Big)^2 \Big(\sin\theta_{12}^{R}\sin\theta_{24}^{R}+\cos\theta_{12}^{R}\cos\theta_{24}^{R}\sin\theta_{14}^{R}\Big)^2 G(x_1) \\
&+ \bigg( \lvert \Big(\sin\theta_{12}^{L}\sin\theta_{24}^{L}+\cos\theta_{12}^{L}\cos\theta_{24}^{L}\sin\theta_{14}^{L}\Big)\Big(\cos\theta_{12}^{L}\sin\theta_{24}^{L}-\cos\theta_{24}^{L}\sin\theta_{12}^{L}\sin\theta_{14}^{L}\Big) \rvert^2 \\
&+ \lvert \Big(\sin\theta_{12}^{R}\sin\theta_{24}^{R}+\cos\theta_{12}^{R}\cos\theta_{24}^{R}\sin\theta_{14}^{R}\Big)\Big(\cos\theta_{12}^{R}\sin\theta_{24}^{R}-\cos\theta_{24}^{R}\sin\theta_{12}^{R}\sin\theta_{14}^{R}\Big) \rvert^2 \bigg) F(x_2) \\
&+ \frac{m_2}{m_1} \Big(\sin\theta_{12}^{L}\sin\theta_{24}^{L}+\cos\theta_{12}^{L}\cos\theta_{24}^{L}\sin\theta_{14}^{L}\Big)\Big(\cos\theta_{12}^{L}\sin\theta_{24}^{L}-\cos\theta_{24}^{L}\sin\theta_{12}^{L}\sin\theta_{14}^{L}\Big) \\
&\times \Big(\sin\theta_{12}^{R}\sin\theta_{24}^{R}+\cos\theta_{12}^{R}\cos\theta_{24}^{R}\sin\theta_{14}^{R}\Big)\Big(\cos\theta_{12}^{R}\sin\theta_{24}^{R}-\cos\theta_{24}^{R}\sin\theta_{12}^{R}\sin\theta_{14}^{R}\Big) G(x_2) \\
&+ \bigg( \lvert \cos\theta_{14}^{L}\cos\theta_{24}^{L}\Big(\sin\theta_{12}^{L}\sin\theta_{24}^{L}+\cos\theta_{12}^{L}\cos\theta_{24}^{L}\sin\theta_{14}^{L}\Big) \rvert^2 \\
&+ \lvert \cos\theta_{14}^{R}\cos\theta_{24}^{R}\Big(\sin\theta_{12}^{R}\sin\theta_{24}^{R}+\cos\theta_{12}^{R}\cos\theta_{24}^{R}\sin\theta_{14}^{R}\Big) \rvert^2 \bigg) F(x_4) \\
&+ \frac{M_4^C}{m_1} \cos\theta_{14}^{L}\cos\theta_{24}^{L}\Big(\sin\theta_{12}^{L}\sin\theta_{24}^{L}+\cos\theta_{12}^{L}\cos\theta_{24}^{L}\sin\theta_{14}^{L}\Big) \\
&\times \cos\theta_{14}^{R}\cos\theta_{24}^{R}\Big(\sin\theta_{12}^{R}\sin\theta_{24}^{R}+\cos\theta_{12}^{R}\cos\theta_{24}^{R}\sin\theta_{14}^{R}\Big)G(x_4)
\label{eqn:expanded_expression_for_electrong2_mixing_angles}
\end{split}
\end{align}

\subsection{Neutrino trident}

The constraint from neutrino trident has a much simpler form compared to the other observables, as it only depends on coupling of the heavy $Z'$ to two muons.
\begin{equation}
\begin{split}
\frac{\left( g_L \right)_{\mu\mu}^2 + \left( g_L \right)_{\mu\mu} \left( g_R \right)_{\mu\mu}}{M_{Z^{\prime}}^2} = &\frac{\Big(\cos\theta_{12}^{L}\sin\theta_{24}^{L}-\cos\theta_{24}^{L}\sin\theta_{12}^{L}\sin\theta_{14}^{L}\Big)^4}{M_{Z^\prime}^2} \\ 
	&+ \frac{\Big(\cos\theta_{12}^{L}\sin\theta_{24}^{L}-\cos\theta_{24}^{L}\sin\theta_{12}^{L}\sin\theta_{14}^{L}\Big)^2 \Big(\cos\theta_{12}^{R}\sin\theta_{24}^{R}-\cos\theta_{24}^{R}\sin\theta_{12}^{R}\sin\theta_{14}^{R}\Big)^2}{M_{Z^{\prime}}^2}
\end{split}
\end{equation}
\end{document}